\RequirePackage{fix-cm}

\documentclass[twocolumn]{svjour3}

\usepackage{graphicx}
\usepackage{xcolor}

\usepackage{harveyballs}

\newcommand{\hbfull}{
    \def\harveyBallsColor{gray}
    \def\harveyBallsLineColor{gray}
    \harveyBallFull{}
}
\newcommand{\hbrquarter}{
    \def\harveyBallsColor{gray}
    \def\harveyBallsLineColor{gray}
    \reflectbox{\harveyBallQuarter}
}

\newcommand{\hbrhalf}{
    \def\harveyBallsColor{gray}
    \def\harveyBallsLineColor{gray}
    \reflectbox{\harveyBallHalf}
}
\newcommand{\hbempty}{
    \def\harveyBallsColor{gray}
    \def\harveyBallsLineColor{gray}
    \harveyBallNone{}
}

\begin{document}

\title{Building Domain-Specific Machine Learning Workflows: A Conceptual Framework for the State-of-the-Practice}

%\subtitle{Finding the Right Level of Low-code for Machine Learning}


\author{Bentley James Oakes         \and
        Michalis Famelis \and
        Houari Sahraoui%etc.
}


\institute{B. Oakes \at
              DIRO, Universit\'e de Montr\'eal, Montreal, Canada \\
              \email{bentley.oakes@umontreal.ca}
            \and M. Famelis
            \at
              DIRO, Universit\'e de Montr\'eal, Montreal, Canada \\
              \email{famelis@iro.umontreal.ca}
           \and
           H. Sahraoui \at
              DIRO, Universit\'e de Montr\'eal, Montreal, Canada \\
              \email{sahraouh@iro.umontreal.ca}
}

%\institute{B. Oakes \at
%              Universit{\'e} de Montr{\'e}al,\\
%              Montr{\'e}al, Canada\\
              %\email{bentley.oakes@umontreal.ca}           %  \\
        %   \and
        %   S. Author \at
        %       second address
%}

\date{}%Received: date / Accepted: date}

\maketitle

\begin{abstract}
Domain experts are increasingly employing machine learning to solve their domain-specific problems. This article presents six key challenges that a domain expert faces in transforming their problem into a computational workflow, and then into an executable implementation. These challenges arise out of our conceptual framework which presents the ``route'' of options that a domain expert may choose to take while developing their solution.

To ground our conceptual framework in the state-of-the-practice, this article discusses a selection of available textual and graphical workflow systems and their support for these six challenges. Case studies from the literature in various domains are also examined to highlight the tools used by the domain experts as well as a classification of the domain-specificity and machine learning usage of their problem, workflow, and implementation.

The state-of-the-practice informs our discussion of the six key challenges, where we identify which challenges are not sufficiently addressed by available tools. We also suggest possible research directions for software engineering researchers to increase the automation of these tools and disseminate best-practice techniques between software engineering and various scientific domains.

\end{abstract}

%\subsubsection*{Machine learning, domain-specific, pipelines, computational workflows, notebooks}
%\keywords{}

\section{Introduction}
\label{sec:intro}

The past two decades have seen the learning algorithms, especially deep learning, permeate throughout every scientific, engineering, and business \emph{domain} to enable new techniques and solve complex challenges. One example of many is the recent solving of the long-standing protein folding problem, which focuses on how a protein will fold in three-dimensional space given it's one-dimensional representation~\cite{Jumper2021}.

The enormous power offered by current machine learning techniques is therefore of great interest to stakeholders across all domains. However, utilising these techniques often requires a user who is an expert in their own domain to gain proficiency in not only the concepts of machine learning but also the programming abilities used to call the low-level libraries. This is undesirable as the \textit{domain expert} would like to reason about concepts and terms that are in their own domain. For example, the literature on \textit{domain-specific languages} shows that solving a problem in the \textit{problem domain} is more efficient than solving it in the \textit{solution domain}~\cite{Kaernae2009,Voelter2019}.

Recently, the rise of \textit{low-code} platforms partially addresses this issue by raising the level of abstraction from writing code to interactively defining procedures and GUIs ~\cite{Cabot2020,Bock2021}. A domain expert (or ``citizen developer'') is thus assisted to build applications or computations using an easy-to-use and easy-to-deploy interface. The domain expert may also be able to select \textit{domain-specific} components which directly address concerns in their domain such as IoT~\cite{Ihirwe2020}, or \textit{machine learning} components to simplify machine learning tasks.

%As a final step, this pipeline must also be executed to perform the calculations. This can involve deployment, code generation, or execution within a pipeline tool.

\subsubsection*{Research Problem}

In this article, we focus on this intersection of \textit{domain experts}, \textit{low-code solutions}, and \textit{machine learning}. Specifically, we are interested in the \textit{flow-based}~\cite{Morrison1994} nature of \textit{workflows}, which are the typical presentation of computations in low-code platforms and \textit{scientific computing frameworks}.
That is, these workflows are composed of computational blocks arranged in a graph structure with the edges denoting data or control dependencies. We note that this representation perfectly matches with the common notion of a machine learning \textit{pipeline} where data is ingested, cleaned, and trained upon to produce a machine learning model.

%This article focuses on providing a conceptual framework to analyse how domain experts can leverage machine learning techniques to solve their problems. In particular, we focus on the representation of computations as \textit{workflows} due to this interesting overlap between low-code, scientific computing, and machine learning use of workflows and pipelines.

The research question that naturally arises is thus:
%studied by this emerging line of research is thus the following:
\textit{Given a domain expert and a domain-specific (DS) problem, what are the tools and techniques needed such that this problem can be: a) mapped to a machine learning (ML) representation, b) constructed as a workflow which utilises ML techniques and libraries, and c) deployed from that workflow into an executable implementation}? 

For instance, to address the first sub-question, we have identified that a domain-specific problem can be mapped to a ML representation either \textit{formal reasoning} or \textit{expert mapping}. Huang \textit{et al.} describe such an expert mapping in the domain of genomics~\cite{Huang2021}. Their work assists genomics researchers by mapping drug and tissue structures from the domain into forms suitable for ML, defining the problem as a ML problem, and recommending suitable ML techniques for solving the ML problem.

% Thus the genomics expert is assisted in building or finding a suitable workflow to solve their problem.

%Thus, Huang \textit{et al.} provide an expert voice to transform the domain-specific problem into a machine learning pipeline representation.

\subsubsection*{Contributions}

Our main contribution is a conceptual framework to map out the tools and techniques for solving a DS problem using a workflow which involves ML, where that workflow is then deployed into an executable implementation. This framework illustrates that there are multiple \textit{choices} or \textit{routes} that a domain expert may choose from to obtain an executable implementation from their problem. For example, a domain expert may wish to first consult an \textit{expert mapping} to determine how the problem should be structured in a ML representation, before they construct a workflow by \textit{manually selecting suitable components from a repository}.

These three layers are divided into \textit{regions}, where \textit{problems}, \textit{workflows}, and \textit{implementations} are organised by the dimensions of \textit{domain-specificity} and \textit{presence of machine learning}. The choices of the domain expert can then be seen as \textit{transformations} between different regions of our conceptual framework. These transformations are phrased as interesting software engineering challenges which directly impact a domain expert who wishes to use machine learning.
%\jessie{Maybe introducing the following challenges can wait Section 3.2 after the definitions of layers, regions and transformations?}
In other words, we attempt to organise a) the meaning of, and b) tools and techniques such that a domain expert can:
\begin{itemize}
    \item Map a DS problem to a form suitable for ML
    \item Obtain a solution workflow for a DS and/or ML problem
    \item Experiment with ML tools and techniques within a workflow
    \item Add DS knowledge to improve ML performance (e.g., feature engineering)
    \item Produce an implementation from a workflow which is well-suited for a domain expert (in terms of scalability, DS tooling, etc.)
    \item Extract a workflow from an existing implementation (code, Jupyter notebook)
\end{itemize}

To ground the above framework and challenges, this article discusses some state-of-the-practice tools and techniques which address these questions. A selection of case studies also provides recent examples of domain experts utilising machine learning as framed in the context of our framework. We also provide a detailed discussion of the benefits of structuring the research problem into our conceptual framework, along with an examination of challenges and research directions.

Overall, our intention with this article is to provide a starting point for those readers who are interested in overcoming these challenges such that domain experts can more efficiently utilise machine learning to solve their problems. This involves cross-discipline efforts to apply best practices and insights across software engineering, machine learning, and other scientific fields.

%ground these choices in a narrative and provide discussion around the benefits and drawbacks of each choice.

\subsubsection*{Article Structure}

As our framework relates topics from different disciplines, Section~\ref{sec:background} provides background on the topics of \textit{domain-specificity}, \textit{machine learning}, and \textit{workflows}.

Our three-layer conceptual framework relating \textit{problems}, \textit{solution workflows}, and \textit{implementations} is presented in an overview in Section~\ref{sec:overview}.  Section~\ref{sec:layers} discusses each layer of this framework in turn including the transformations within each layer. Section~\ref{sec:transformations} then presents the transformations between layers. These inter-layer transformations involve turning problems into workflows, and from workflows into an implementation.

Section~\ref{sec:tools} provides a brief selection of state-of-the-practice tools which implement the transformations found in our framework. Representative case studies are presented in Section~\ref{sec:case_studies} to detail how practitioners from various fields are employing these tools and executing these transformations on their own problems.

Section~\ref{sec:discussion} then uses the framework as a basis for a discussion of the opportunities and challenges for implementing and automating the transformations in the framework. Section~\ref{sec:conclusion} presents our concluding thoughts.

\section{Background}
\label{sec:background}

This section provides a brief background in three core topics of this article: the concept of \textit{domain-specificity}, \textit{machine learning}, and \textit{computational workflows}.

\subsection{Domain-Specificity}

In this article, we discuss the idea of \textit{domain-specificity} as relating to the concepts of a particular domain. This is relevant throughout our framework as a) a domain expert had specific experience and insights into that domain and is less familiar with concepts outside that domain, b) the problem the domain expert has is (partially) expressed using those domain concepts, and c) the technical considerations for a solution such as computing platforms and tools may be specific to that domain. In this article, \textit{domain-specificity} is manifested as a cross-cutting \textit{dimension} of the framework as discussed in Section~\ref{sec:dimensions}. In particular, we specify that \textit{problems}, \textit{workflows}, and \textit{implementations} can be more or less \textit{domain-specific}.

For example, consider the domain of \textit{neuroscience} which is the study of the nervous system. Relevant concepts to a neuroscientist include \textit{neurons}, \textit{signals}, \textit{behaviour}, \textit{activation}, \textit{brain hemispheres}, \textit{neural circuits}, and \textit{brain damage}, with more according to the sub-field of the neuroscientist. These concepts may be formalised in an \textit{ontology}\footnote{See \url{https://github.com/SciCrunch/NIF-Ontology} for an example.}, allowing for more precise or (semi-) automated reasoning about the concepts and their connections.

Along with these concepts, the data examined by this expert is highly specific to the domain. In neuroscience, this can include processing of functional Magnetic Resonance Imaging (fMRI) files requiring specialised techniques to handle \textit{motion correction} (correcting the movement of the subject within the scanner) and \textit{smoothing} of the data (to average out the noise present in the measurement). Finally, the datasets, tools, and computational platforms available to a neuroscientist are highly domain-specific such as repositories of mouse brain scans\footnote{For example: \url{https://neuinfo.org/}.}.

As a domain expert is most knowledgeable about concepts from their domains, we follow the approach of \textit{domain-specific modelling} in declaring that problems should be solved at a high level of abstraction using domain-specific concepts~\cite{Kaernae2009,Voelter2019}. That is, the domain expert should use \textit{domain-specific languages} instead of general programming languages, and there should be a layered approach when possible to hide technical and implementation details. This approach has been shown to lower the cognitive workload of learning new concepts and resulting in increased productivity~\cite{Voelter2019}. In the context of this article, we are interested in improving support such that the domain expert can describe their problem and workflow using domain-specific concepts.

%Inputs, outputs, concepts related to a domain. From biology, programming (CICD/DevOps), materials science, sociology, etc.

%Syntax and semantics of concepts, including restrictions on the entered structure or data.

%Could also include machine learning as a domain itself. For example, a problem in the machine learning domain could be predicting whether a set of parameters will be good or not.

\subsection{Machine Learning}

Machine learning (ML) can be summarised as ``programming computers to optimise a performance criterion using example data or past experience''~\cite{Alpaydin2020}. That is, based on a collection of data and experiences, ML seeks to automatically create \textit{models} capable of relating output for particular inputs without requiring a programmer to directly implement the steps for computing such outputs. These definitions therefore capture many applications ranging from interpolating based on a simple linear regression up to automated driving by using visual data interpreted using neural networks.

One way of categorising ML approaches is into three approaches: \textit{supervised learning}, \textit{unsupervised learning}, and \textit{reinforcement learning}. In \textit{supervised learning}, the supervisor provides inputs and labelled outputs, and the technique must learn the mapping between inputs and outputs. An example would be to train a linear regression of variables, producing a classifier which can predict whether a tissue sample is a tumor or not~\cite{Foell2021}, or constructing a similarity function such that related objects can be found. In \textit{unsupervised learning}, the technique learns without provided labels. This can offer insights into the structure of the domain such as uncovering related clusters of data or outliers. In \textit{reinforcement learning}, the intelligent agent is rewarded with a defined reward function. This allows the agent to explore possible actions and receive automated feedback.

A ML model must be \textit{trained} before it can be used for \textit{reasoning}. For example, consider the scenario of training a model for a ``line-of-best-fit'' (linear regression). Once the data points have been loaded and any cleaning necessary performed, the data is then commonly divided into \textit{training} and \textit{test} sets. This allows for an unbiased measure of error when testing, as the model has not ``seen'' the testing data beforehand. The linear regression is then learned by a suitable algorithm, and the linear regression is the produced ML model ready for use. When this model is used it is fed a new piece of data as input, and it will predict a particular output.

%This separation of ML model \textit{training} and \textit{reasoning} produces the notion of model \textit{producers} and model \textit{consumers}.

Many libraries are available which implement some form of ML. For example, scikit-learn\footnote{\url{https://scikit-learn.org}} is a popular Python module which offers a high-level API for ML techniques, while Keras\footnote{\url{https://keras.io}} provides an API for directly creating neural networks by constructing each layer into a model. ML techniques are also available directly within academic and scientific tools such as the MATLAB Statistics and Machine Learning toolbox\footnote{\url{https://www.mathworks.com/products/statistics.html}}.

\subsection{Computational Workflows}
\label{sec:background_workflows}

General workflows have existed for many years, especially in the manufacturing domain. In this article, we focus on \textit{computational workflows} which define the steps that a computational device follows to produce results.

In particular, we are interested in workflows which represent \textit{flow-based programming} by containing discrete steps representing computational steps~\cite{Morrison1994}. This can be represented by a directed-acyclic graph (DAG) of computational nodes connected by control and data dependencies. Note that in current workflow systems (Section~\ref{sec:tools}) this restriction on acyclic graphs is relaxed, as tools may wish to encode control flow structures such as loops over input files.

This concept of data ``flowing'' through a workflow is quite a natural structure for many computations we examine in this article. Indeed, some domains may lean into the plumbing metaphor and refer to the workflow as a ``pipeline''. Lamprecht \textit{et al.} state that, ``Another common, more differentiating view is that pipelines are purely computational and as such a subset of the more general notion of workflows, which can also involve a human element"~\cite{Lamprecht2021}. To clarify the terminology in this article, we only discuss \textit{computational workflows}. That is, the term `workflow' in this article do not contain human elements and can therefore be equated to \textit{pipelines}.

%Despite the terminology difference, what's important is that there is a directed graph with computation nodes.

\subsubsection*{Scientific Workflow Management}

The large scale of scientific data and the \textit{reproducibility} requirements of scientific processes have encouraged the growth of \textit{workflow management systems} within various scientific domains. For example, ensuring that data sources and computations follow the well-known principles of \textit{Findable}, \textit{Accessible}, \textit{Interoperable}, \textit{Reusable} (FAIR)~\cite{Wilkinson2016} demands a comprehensive management system. The life sciences in particular have a rich history of workflow systems~\cite{Leipzig2017,Wratten2021,Paul-Gilloteaux2021,Lamprecht2021}, a few of which are discussed in our sections on tools (Section~\ref{sec:tools}) and case studies (Section~\ref{sec:case_studies}).

As workflows explicitly declare the sequence of computations they perform, they assist in the production of \textit{reproducible} results~\cite{Theaud2020,Wratten2021}. That is, they assist to reproduce the results of another experiment~\cite{Ivie2018}. For example, the discrete nature of workflow components means that components can be tagged with provenance information~\cite{Salazar2021} and placed into containers such as Docker containers~\footnote{\url{https://www.docker.com}}. This allows for easily accessible, self-contained units which can be accessed through repositories and placed into dependency management systems~\cite{DiTommaso2017}.

Many factors can influence whether a computational process is reproducible, and workflows are only one step towards full reproducibility. For example, Digan \textit{et al.} discuss 40 reproducibility features sourced from recommendations and clinical usages of workflows in the natural language processing (NLP) domain~\cite{Digan2021}. Mora-Cantallops \textit{et al.} discuss reproduciblity in the context of artifical intelligence/ML~\cite{Mora-Cantallops2021}.

% Theaud2020 develops reprocudable/replicable workflow for diffusion MRI tractography.
% \url{https://github.com/scilus/tractoflow} Looks at what parameters affect reproducibility.

\subsubsection*{Machine Learning Pipelines}

In ML, the term \textit{pipeline} is commonly used to denote the steps involved in the \textit{training} and \textit{reasoning} processes. For example, data can be involved in a linear flow of \textit{loading}, \textit{filtering}, \textit{cleaning}, \textit{splitting} (into a training and testing set), and \textit{trained upon}.

A similar linear flow is also present for the neural networks used in \textit{deep learning}. The input data passes through \textit{layers} in this network which recognise patterns in the input data, store relationships in the weights between nodes in inner layers, and then produce output values or categories.

Thus, ML pipelines are a one-to-one match to the workflow formalism we examine in this article. This unification of structure is important when we discuss workflows which involve components from both a particular scientific domain as well as ML components.
\section{Overview of Our Framework}
\label{sec:overview}

This section will provide an overview of our conceptual framework by discussing its structure, two dimensions, and relating the framework to our challenge questions introduced in Section~\ref{sec:intro}.

\begin{figure*}[tbh]
    \centering
    \includegraphics[width=0.8\textwidth]{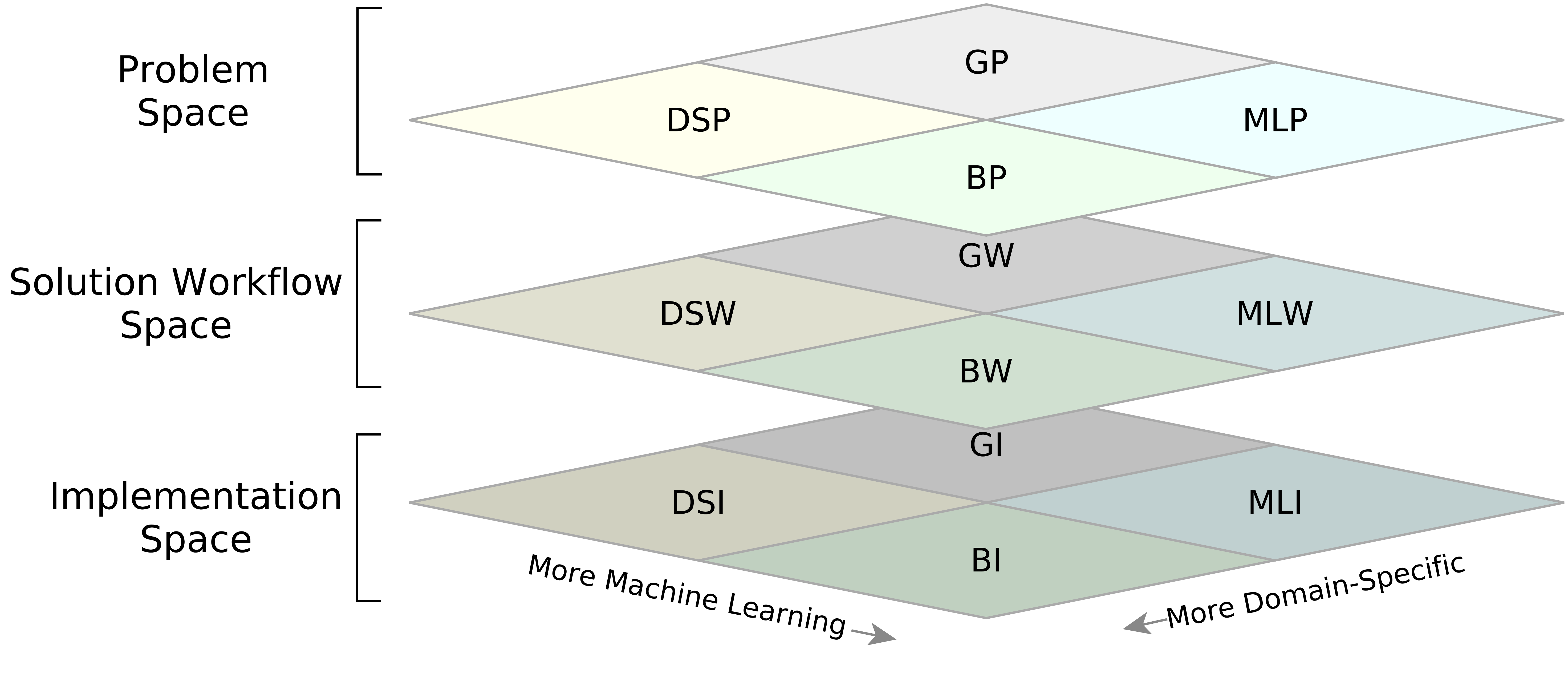}
    \caption{Our three-layer framework with the \textit{problem}, \textit{solution workflow}, and \textit{implementation} spaces.}
    \label{fig:overview}
\end{figure*}

As a summary, our conceptual framework maps out the options for a domain expert to choose from to develop a workflow-based solution to their problem using machine learning, and ultimately obtain a usable implementation. The framework is seen in Figure~\ref{fig:overview}, separated into twelve regions through the division of three layers and two dimensions. The three horizontal layers define the \textit{problem space}, \textit{workflow solution space}, and the \textit{implementation space}. The \textit{problem space} layer contains the \textit{problem} of the domain expert. For example, a neuroscientist may wish to classify brain scan data as belonging to a depressed patient or not. The \textit{solution workflow space} layer contains workflows which solve the problems from the upper layer. The \textit{implementation space} contains the implementation details for the workflows in the middle layer. Section~\ref{sec:layers} addresses these layers in detail.

Each layer is decomposed into four regions, defined by the \textit{domain-specific} and \textit{machine learning} dimensions. Section~\ref{sec:layers} also discusses the mappings and transformations between the regions in a layer. For example, one transformation would map a problem in the \textit{domain-specific problem space} (DSP) to the \textit{machine learning problem space} (MLP). This represents all techniques to transform the \textit{domain-specific problem} to a \textit{machine learning problem} such as an \textit{expert mapping}~\cite{Huang2021}.

Section~\ref{sec:transformations} will then discuss the mappings and transformations \textit{between} layers. That is, from a problem to a workflow, and then from a workflow to an implementation.

\subsection{Dimensions of the Space}
\label{sec:dimensions}

This framework defines two dimensions for each layer. First is the notion of \textit{domain specificity} which captures how many domain-specific concepts an artefact contains. Second, the \textit{machine learning dimension} captures how many machine learning (ML) concepts the artefact contains.

For example, consider an artefact from the first layer: a \textit{problem} to be solved. This problem could be very generic such as ``\textit{finding the clusters for a table of data}''. Adding \textit{domain specificity} comes from adding domain knowledge to the problem, such as \textit{clustering related genes or molecules} which may change the solutions available for the problem. Similarly, ML concepts such as \textit{clustering}, \textit{unsupervised/supervised learning}, or \textit{convolutional filters} may be present in the problem description or not, affecting how far the problem is along the \textit{machine learning} dimension.

In the \textit{problem space} layer, the problems are categorised based on the domain-specific or ML concepts present. For the \textit{solution workflow} layer, this categorisation depends on the proportion of components in the workflow. In the \textit{implementation space} layer, this categorisation depends on the use of domain-specific or ML APIs or libraries. Section~\ref{sec:layers} further discusses how the dimensions divide up each layer in the framework. 

Note that by necessity, these categorisations are \textit{fuzzy} and we intentionally do not provide boundaries to these regions. Instead, we provide these dimensions to provoke reflection about the \textit{mappings} and \textit{transformations} along these dimensions. That is, what does it mean to make a problem, workflow, or implementation \textit{more domain-specific} or involve \textit{more machine learning}, and what are the techniques to do so?

% We also wish to connect the defined regions to state-of-the-practice tools and case studies in Section~\ref{sec:tools} and Section~\ref{sec:case_studies}. That is, how the case studies use various tools to transform their artefacts through the framework.

Let us also note that these two dimensions are certainly not orthogonal. Specifically, \textit{machine learning} is itself a domain of interest and these two dimensions may overlap significantly depending on the problem of interest. Therefore, let us state here that when this article refers to \textit{domains} or \textit{domain experts}, we refer to a non-machine learning domain.

\subsection{Relation to Software Engineering Challenges}

As expressed in our framework, the domain expert wishes to transform their problem from their domain (the DSP region on the top layer in Figure~\ref{fig:overview}) all the way into an implementation using machine learning libraries (the BI or MLI regions on the bottom layer). Through these transformations, the domain expert is able to move around through different regions as shown by the case studies in Section~\ref{sec:case_studies}, though support may be lacking for some operations in various tools as discussed in Section~\ref{sec:discussion}. In particular, we are interested in determining tool support for transformations enabling the most direct route from the domain-specific problem (DSP) to a blended workflow (BW) down to a blended implementation (BI).

Here we recall the challenge questions from Section~\ref{sec:intro} and relate them to specific transformations. Note that other transformations are indeed possible and may be relevant to a domain expert. However, we have chosen these transformations to focus on as they seem most relevant to the tools and case studies discussed in this article.

%\bentley{Keep these in sync.}

\paragraph{Mapping a DS problem to a form suitable for ML}:
This challenge refers to how domain-specific problems in the \textit{problem space} can be mapped or transformed to include more ML concepts. That is, moving the problems along the \textit{ML dimension} in the \textit{problem space}.

\paragraph{Providing a solution workflow for a DS and/or ML problem}:
This challenge refers to the mapping between a problem in the \textit{problem space} and a workflow in the \textit{solution workflow space}. That is, moving from the top to the middle layer in the workflow.

\paragraph{Allowing the domain expert to experiment with appropriate ML components in a DS workflow}:
This challenge refers to moving solution workflows on the middle layer along the ML dimension through the integration of more ML components.

\paragraph{Adding DS knowledge to improve ML performance}:
This challenge refers to moving solution workflows on the middle layer along the DS dimension by adding new DS information or components.

\paragraph{Producing an implementation from a workflow which is well-suited for a domain expert}:
This transformation is between the middle and bottom layers, as the workflow of the domain expert is mapped or transformed to an executable implementation.

\paragraph{Extracting a workflow from an existing implementation (code or notebook}:
This challenge is an inverse to the previous one, as the implementation on the bottom layer of the framework is instead transformed into a solution workflow on the middle layer.

% Thus the domain expert has to choose which way to go, based on the tools available. Software engineer's role is to provide tools and techniques.

\section{Layers and Intra-Layer Transformations}
\label{sec:layers}

This section details the three layers of our framework as shown in Figure~\ref{sec:overview}: the \textit{problem space}, the \textit{solution workflow space}, and the \textit{implementation space}. The \textit{regions} and relevant \textit{transformations} within each layer are then presented.

\subsection{Problem Space}
\label{sec:problem_space}

\begin{figure}[tbh]
    \centering
    \includegraphics[width=0.48\textwidth]{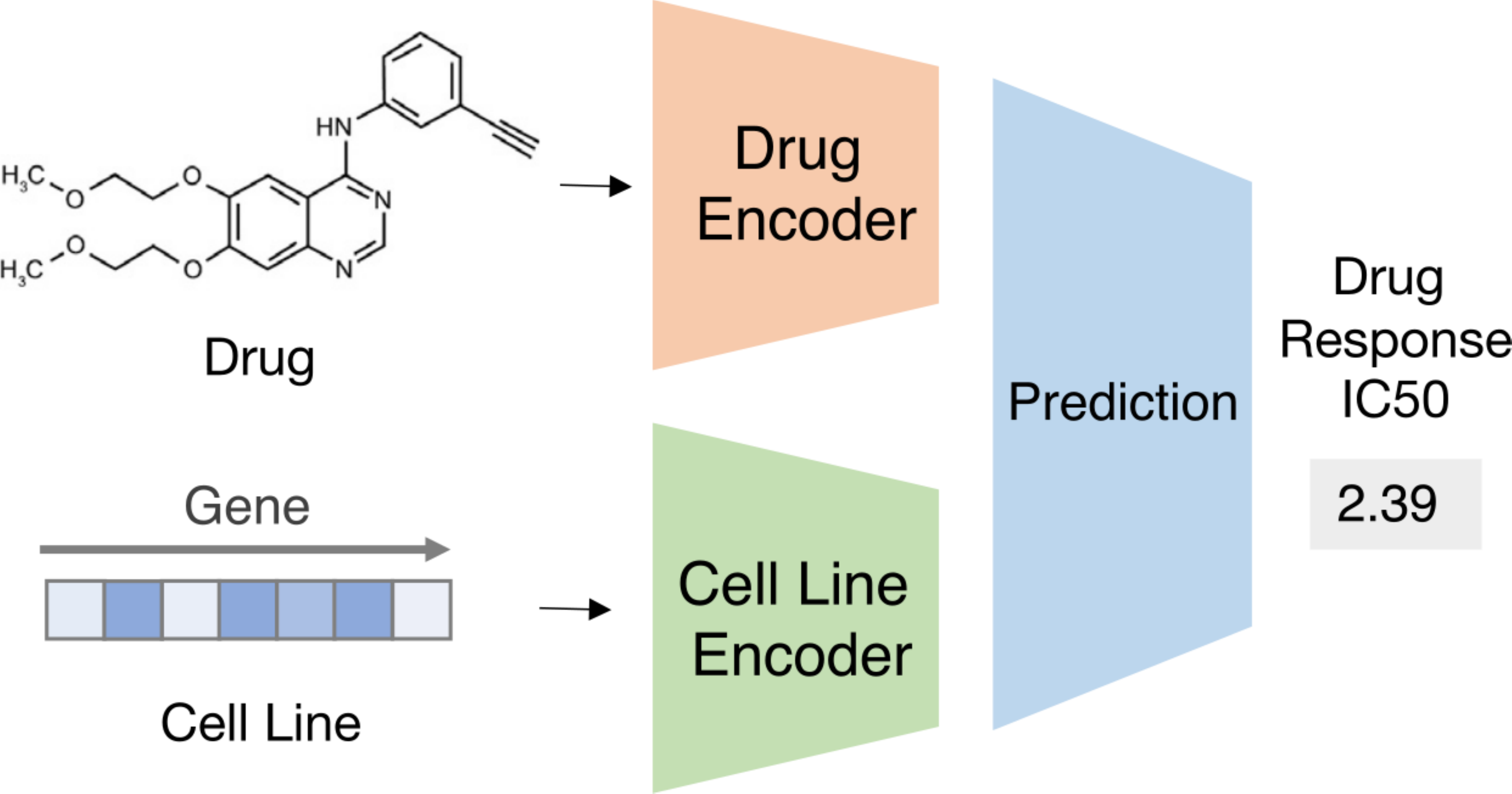}
    \caption{The \textit{Drug Response Prediction} problem from Huang \textit{et al.}~\cite{Huang2021}.}
    \label{fig:huang_drug}
\end{figure}

The \textit{problem space} is where the problem of the domain expert is formulated. For example, the running example presented in Section~\ref{sec:intro} can be expressed in two ways. In the \textit{domain-specific} space (DSP), the problem is to predict whether a drug will work well with a sample of tissue~\cite{Huang2021}. As a \textit{machine learning} representation (MLP), this problem becomes a numerical prediction for efficacy of the graph structure (for the drug) versus a linear string of characters (for the genes present in the tissue). A \textit{blended} version of this problem (BP) is visualised in Figure~\ref{fig:huang_drug} where the structures for the drug and gene are fed into ML encoders and used to make a numerical prediction related to how the drug affects the gene.

\subsubsection{Artefact Representation}

The artefact for this layer is a \textit{problem} composed of \textit{concepts}. This problem may be specified in multiple ways ranging from a simple informal statement to a complex formal representation. For example, the problem could be specified as natural language research questions, in an ontological manner, in a textual or graphical domain-specific language (DSL), or in a template-based, requirement-like manner.

\subsubsection{Layer Regions}

\begin{figure}[tbh]
    \centering
    \includegraphics[width=0.48\textwidth]{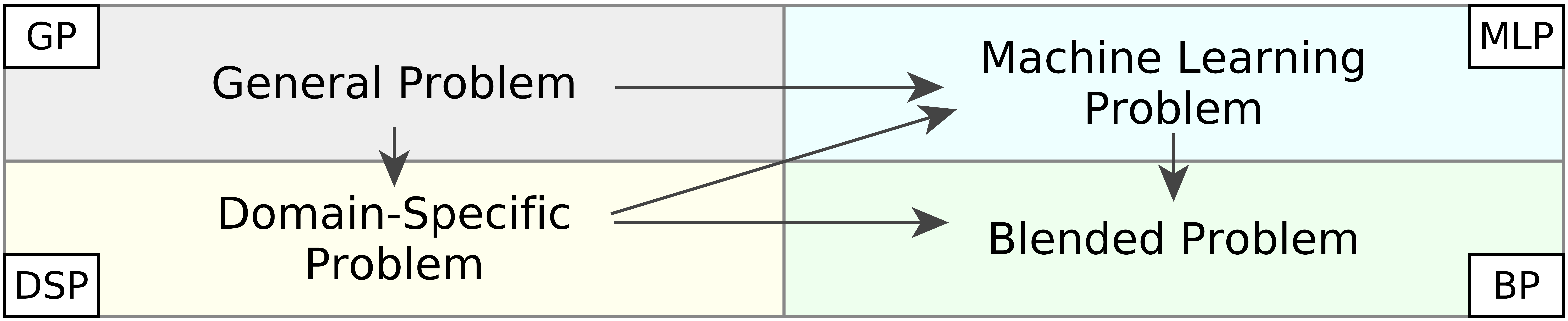}
    \caption{A representation of the problem space, with intra-layer transformations.}
    \label{fig:problem_space}
\end{figure}

In our framework, we define four \textit{regions} for the problem space: \textit{General Problems} (GP), \textit{Domain-Specific Problems} (DSP), \textit{Machine Learning Problems} (MLP), and \textit{Blended Problems} (BP) as shown in Figure~\ref{fig:problem_space}. These regions are defined by the \textit{domain-specificity} and \textit{machine learning} dimensions as discussed in Section~\ref{sec:dimensions}.

A \textit{general problem} is one which does not refer to any domain-specific concepts, nor does it refer to any machine learning concepts. This region is provided for completeness as this article focuses on the other three regions.

\textit{Domain-specific problems} are those that are specified in the domain of interest by experts. They involve concepts which are specialised to that domain. In this article, the domain-specific problem is the starting point from which the expert wishes to solve with a workflow and finally an executable implementation.

A \textit{machine learning problem} is a problem which involves a high proportion of machine learning concepts. For example, learning how to classify objects by ingesting data from a table. The machine learning problem may also delve into high-level workflow steps, such as referring to concepts such as convolutional layers or optimising of meta-parameters in training.

A \textit{blended problem} is one that contains both domain-specific and machine learning concepts. For example, the problem represented in Figure~\ref{fig:huang_drug} contains both DS and ML concepts. This type of problem may be ideal for the domain expert to begin with instead of a domain-specific one, as the ML concepts can be directly conceptualised and/or operationalised in the workflow layer. However, it already implies that the practitioners has access to both domain and machine learning knowledge, and has studied the problem from both domains. %Thus, we focus this article on starting from the \textit{domain-specific problem}.

\subsubsection{Layer Transformations}
\label{sec:layers_pl_trans}

As discussed in Section~\ref{sec:dimensions}, it is interesting to reason about how to transform an artefact on each layer to make it more domain specific or involve more machine learning components. In other words, to make these problems \textit{more DS/ML specific and less general}.

To \textit{increase domain specificity}, a domain expert will have to encode more domain knowledge into the problem. This could be in the form of an ontology representing formal knowledge, or by directly specifying features of interest in the domain and their interaction with other features. For example, consider the drug molecule and gene running example. As this molecule is a chemical structure, this restricts the possible forms it may take, which may make the solving of this problem easier or enable the use of domain-specific tools for the workflow. 

Increasing the \textit{proportion of ML concepts} is similar. A ML domain expert will identify structures and techniques from the ML domain to cast the problem into. This may be to specify the technique used for classification or learning, or the representation that the problem should take. This is the transformation which represents our challenge question: \textit{mapping a DS problem to a form suitable for ML}.

\paragraph{Direct Mapping}
Transformations are also possible directly between the domain-specific and machine learning problem regions.
This transformation is how to take a problem which exists in the domain and map it to a problem in the machine learning space, or to create a many-to-many mapping. This opens up further possibilities for insights and implementation.

Here we identify two possible techniques for mapping: expert mapping and ontological/formal mapping.

In \textit{expert mapping}, a group of experts from both the domain of interest and the machine learning domain issue recommendations about how to map the problem. This is shown by Huang \textit{et al.} where problems in genomics are mapped to machine learning representations~\cite{Huang2021}. This is a high effort technique, but could greatly assist with workflow and implementation creation.

This expert mapping is of course possible in other domains. Aneja \textit{et al.} provide a table mapping the problems addressed in the neuro-oncology literature with the ML approach used in those papers~\cite{Aneja2019}. Jablonka \textit{et al.} provide another detailed review with mention of how problems in material science were mapped to ML~\cite{Jablonka2020}.

Another technique is to perform \textit{formal or ontological reasoning}. In this technique, the two domains would be modelled and a reasoner would perform the mapping. An example would be performing ontological reasoning. This may offer more potential for automation. However, the domains themselves would have to be explicitly modelled~\cite{Soto2021}.

%Also there is lots of work in AutoML, where a framework will start generating solutions from the data itself. This may be applicable here. TO INVESTIGATE.

\paragraph{Reversing the Transformation}

In principle, these transformations could be reversed. That is, a problem could be made \textit{more general} with the same techniques, and a \textit{machine learning} problem could be mapped to a \textit{domain-specific} one. Improving the generalisation and mapping possibilities for problems would increase the workflow and implementations options available to practitioners.

%However, we do not expand on these transformation directions here.

\subsection{Solution Workflow Space}
\label{sec:workflow_space}

In the second layer of our framework, we define a space of \textit{solution workflows}. These workflows detail the actions required to solve the problem defined on the upper layer. 

\subsubsection{Artefact Representation}

The representation of these workflows is based on those described in Section~\ref{sec:background_workflows} with some available standards presented in Section~\ref{sec:tools_workflow_standards}. The core aspect is that these workflows consist of a directed sequence of components with explicit control and/or data flow. These components represent a step or action in the workflow, and they can be typed to enforce structure on the accepted inputs and outputs and define their semantics. 
%\jessie{It could be nice to recall here that a component represents a step/action having inputs/outputs. It is not a programming component, and the workflow's goal is to formulate steps to answer the problem without operationalizing it?}

\begin{figure*}[tbh]
    \centering
    \includegraphics[width=0.8\textwidth]{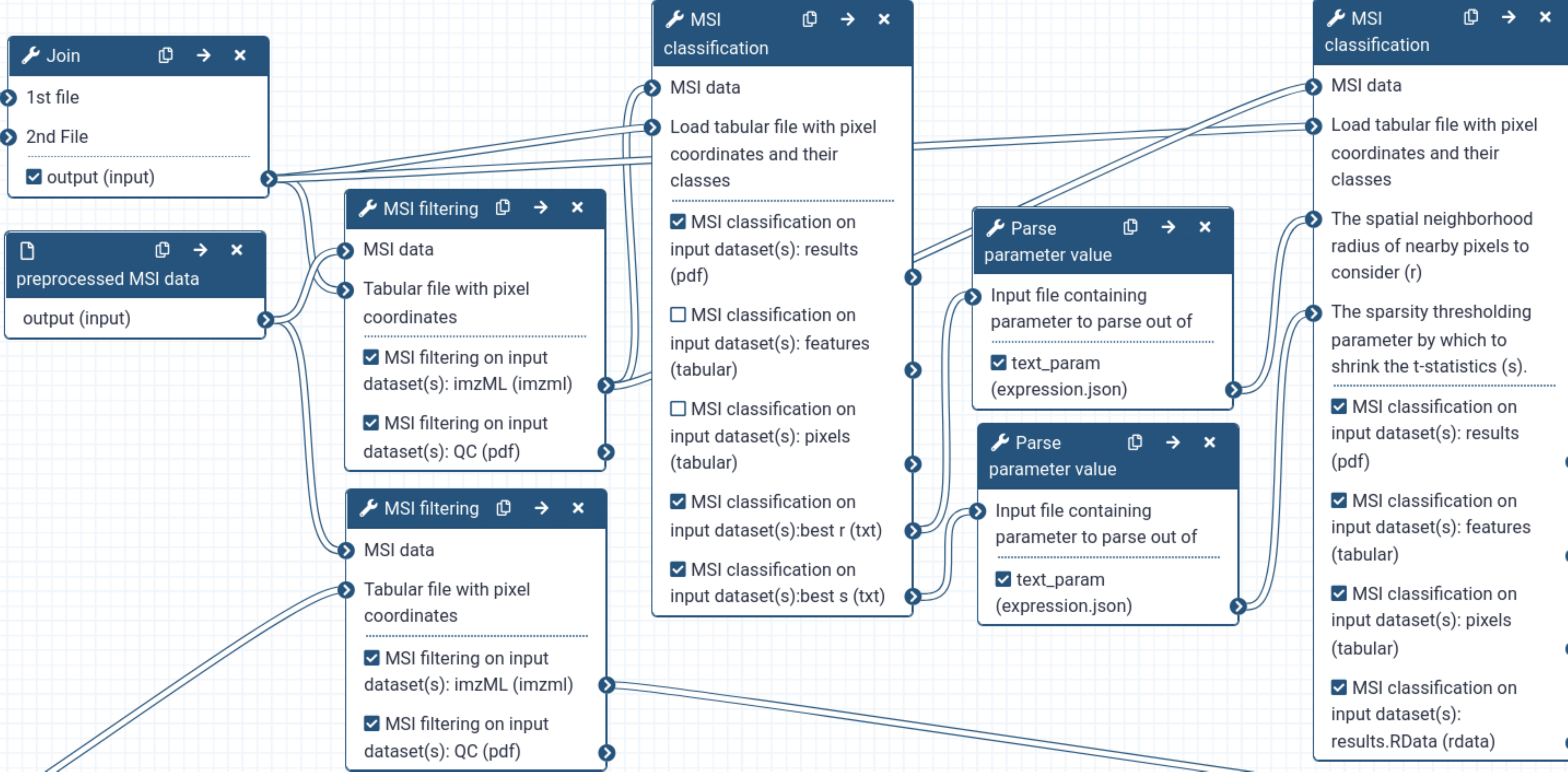}
    \caption{A selection of a workflow described in~\cite{Foell2021}, in the workflow management tool Galaxy~\cite{Jalili2020}.}
    \label{fig:galaxy_workflow}
\end{figure*}

Figure~\ref{fig:galaxy_workflow} shows a few components from a workflow defined in the Galaxy workflow management tool~\cite{Jalili2020}. Galaxy is further discussed in both Section~\ref{sec:tools_graphical} and Section~\ref{sec:cs_galaxy}.

The idea with this representation is to have a unified, graph-like structure amenable for modularity, re-use, modification, and sharing. As mentioned in Section~\ref{sec:background_workflows}, workflows are already common in scientific domains and structured ``pipelines'' are in place in the data science/machine learning world. Therefore our conceptual framework solely focuses on graph-like workflows as the representation for this middle layer.

\subsubsection{Layer Regions}

\begin{figure}[tbh]
    \centering
    \includegraphics[width=0.49\textwidth]{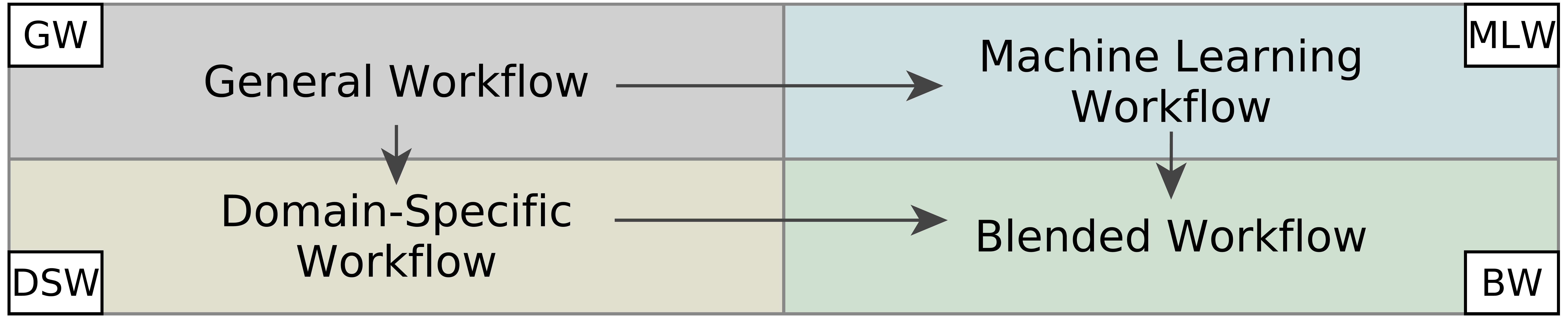}
    \caption{A representation of the solution workflow space, with intra-layer transformations.}
    \label{fig:solution_space}
\end{figure}

In this space, we again roughly classify workflows into four regions as shown in Figure~\ref{fig:solution_space}.

A \textit{general workflow} is one where there are very few domain-specific or machine learning components within the workflow. Thus it is a catch-all category comprising those workflows not discussed below.

In our framework, a \textit{domain-specific workflow} is one that has many components from the domain of interest. These components must address a domain-specific concept which is not of interest to a general user outside of that domain.  Examples include loading a file or database with a domain-accepted structure, a computation performing a specific task of interest to the domain, or communication with domain entities such as robotic arms.

%\jessie{Thus, can we say that a DS component can be defined as an action which is not useful outside the domain of interest? Which cannot be reused outside this domain? I am trying to find some definitions because I have trouble drawing a clear distinction between DS and ML components. I am tempted to think that DS components can always be 'rephrased' as ML components, and thus how could we consider a DS component not being an ML component?}

We also consider a rough spectrum where some workflows have more domain-specific components than others. For example, one workflow may simply load data from an EEG file and visualise it, while another may perform filtering or spectrum analysis on the data. This second workflow is thus more domain-specific than the first.

Similar to our category of domain-specific workflows, a \textit{machine learning workflow} is one that contains ML components. Again, we state that there is a rough spectrum of these workflows from not including machine learning components to heavily relying on these components.

The \textit{blended workflow} is one that contains numerous domain-specific and machine learning components. As well, a component itself may be labelled as \textit{blended}, as it may address a domain specific concern but heavily utilise machine learning ``within'' the execution of the component. We propose that this blended type of workflow is desired for a domain expert to arrive at, since the presence of these components should raise the level of abstraction and increase the modularity and reuse of the workflow. However, it may be preferable for a domain expert not familiar with ML to begin working with a DS workflow, and have the workflow `adjusted' to become more blended over time as they gain familiarity with ML concepts and components. This offers the domain expert further experimentation, customisation, and optimisation possibilities.

\subsubsection{Layer Transformations}

We have posed two interesting questions in the introduction of this article of how to \textit{transform} a workflow along these two dimensions. That is, to identify the techniques to: a) increase the \textit{domain-specificity}, and/or b) \textit{usage of machine learning} in a workflow. These correspond to our challenge questions of: a) \textit{adding DS knowledge to improve ML performance}, and b) \textit{experimenting with ML tools and techniques within a workflow}.

A first approach is to ask a domain or machine learning expert to study the workflow and identify where components should be added or improved. Their knowledge could then be modelled and implemented for automated approaches to detect and suggest components for use, such as that implemented by Kumar \textit{et al.}~\cite{Kumar2021}. These recommender systems could therefore shift the workflow along the \textit{domain-specific} or \textit{machine learning} dimensions.

For example, a workflow may load genomic data from a table provided in a spreadsheet for further processing. Depending on the requirements of the user, the loading components may be better replaced with a component which is able to download up-to-date genomic data directly from a cloud repository as is available in tools like Galaxy.

Domain-specific workflows can also be utilised as a sub-workflow. For example, Sections~\ref{sec:tools_text} and \ref{sec:cs_nipype} discuss the \textit{fMRIPrep} workflow which is for performing specific pre-processing for neuroscience data. Thus, a domain expert's workflow becomes more DS when fMRIPrep is used.

The error of ML techniques may also be lowered when DS information is used. One aspect of this is the field of \textit{feature engineering} where new data is extracted from the old to reduce the error of machine learning algorithms such as deep learning~\cite{Borghesi2020}. For example, Fan \textit{et al.} study the problem of marking reported software bugs as `valid' or `invalid'~\cite{Fan2018}. From the bug data, they extract new features such as \textit{recent number of bugs by reporter}, \textit{does the bug have a code patch attached}, and \textit{bug text readability scores}.

\paragraph{Reversing the Transformation}

Just as with the layer transformations for the problem space (Section~\ref{sec:layers_pl_trans}), these transformations could indeed be reversed to increase the generality of the workflows. That is, to make a workflow \textit{less} DS and involve \textit{fewer} ML components. This may help to increase the applicability of the workflow across domains, however we do not consider it further.

\subsection{Implementation Space}
\label{sec:implementation_space}

Similar to the above layers, the \textit{implementation space} is also a rough categorisation of possibilities. In our conceptual framework, the implementation space defines the low-level result which runs on a computational device, such as produced code or the execution of a workflow engine. For example, some workflow management tools are directly implemented in Python, or the Orange editor (Section~\ref{sec:tools}) can directly execute the workflow inside the tool.

\subsubsection{Artefact Representation}

In this space, the artefacts are represented as code or another machine executable format. This may be Python code which contains calls to ML APIs, or may be the machine code executed by a workflow engine which is directly interpreting and executing the instructions within a workflow.

These two examples are not at the same level of abstraction. However, the intention with this representation is that: a) this code directly calls upon domain-specific and machine learning APIs, and b) this code should not be written directly by a domain expert as the level of abstraction is too low.

\begin{figure}[tbh]
    \centering
    \includegraphics[width=0.49\textwidth]{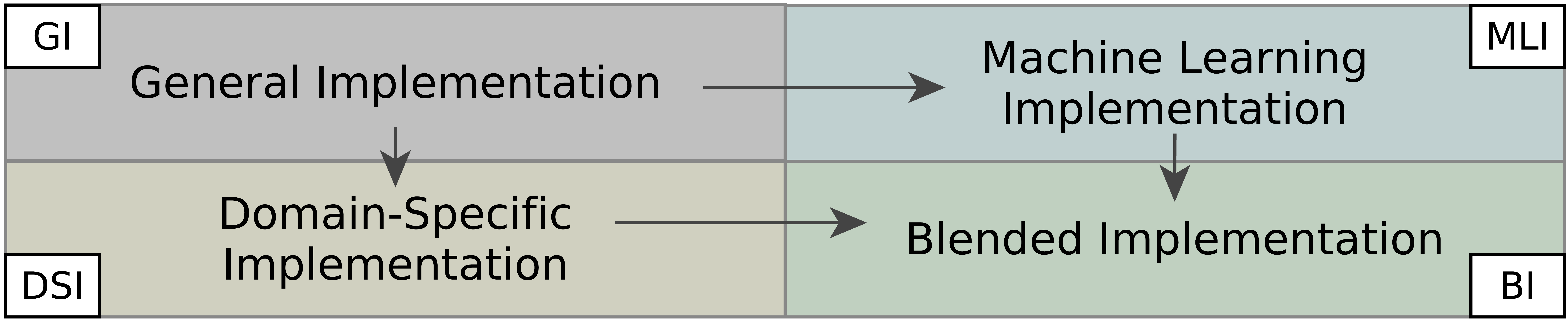}
    \caption{A representation of the implementation space, with intra-layer transformations.}
    \label{fig:implem_space}
\end{figure}

\subsubsection{Layer Regions}

Similar to the above layers, the implementation space layer is defined by the same two dimensions. The \textit{domain-specific} dimension defines the proportion of the code which calls upon domain-specific APIs or libraries. The \textit{machine learning} dimension defines the proportion for machine learning APIs or libraries.

A \textit{general implementation} contains a small proportion of calls to DS or ML APIs or libraries. Thus it is general code.

In a \textit{domain-specific implementation} the code makes calls into an API or library which provides domain-specific computation. For example, the nipype library (Section~\ref{sec:tools}) offers a neuroscience-specific Python library. Thus the more calls to libraries like these, the more domain-specific the implementation.

Likewise, a \textit{machine learning} implementation has a high proportion of machine learning API or library calls. An example would be directly calling Tensorflow or other ML library from Python.

A \textit{blended implementation} is one which uses both \textit{domain-specific} and \textit{machine learning} APIs and libraries. Thus the ML learning libraries are wrapped in a domain-specific interface, and potentially optimised for each domain-specific task. Clearly it would be desirable for the domain expert to produce this form of implementation to achieve the most specific implementation. However, the domain expert should not write the implementation by hand, and instead generate an implementation from their workflow. 

%As we assume that the domain expert is less familiar with ML, the implementation of their solution should be made \textit{as domain-specific as possible}. For complex problems, an initial implementation may be created using ML %

%\jessie{The expert should seek an implementation that uses ML libraries only if a DS library does not provide the needed functionalities? Thus, it is highly unlikely to be able to create a full DS implementation, especially for complex problems?}

\subsubsection{Layer Transformations}

Again, the same transformations as the solution workflow layer occur in this layer. Code can be mapped manually or automatically to either a domain-specific library call or a machine learning one.

These tasks could be useful for addressing legacy code and updating it (called ``cognification''). However, as argued in Section~\ref{sec:discussion}, it is not be efficient to focus on mapping and recommendations at the implementation level. Instead, a more efficient approach would be to extract the workflow from the code and apply analyses at the workflow level.

\section{Inter-layer Transformations}
\label{sec:transformations}

This section defines the possible \textit{transformations} used to transform an artefact in one of the layers in our framework to another layer.  Section~\ref{sec:tools} then provides examples of state-of-the-practice tools which support these transformations.

%These transformations make the artefact more concrete and closer to being executable.

The transformations discussed here are: a) from the \textit{problem space} to the \textit{solution workflow space}, and b) from the \textit{solution workflow space} to the \textit{implementation space}. Note that transforming a problem from the \textit{problem space} to the \textit{implementation space} is \textit{classical programming}, which we do not elaborate further in this section but appears in case studies in Section~\ref{sec:case_studies}.

\subsection{Problem Space to Solution Workflow Space Transformations}
\label{sec:p_space_to_sw_space}

The transformation between the first two layers of our framework transforms \textit{problems} from the \textit{problem space} into \textit{workflows} in the \textit{workflow space}. This transformation corresponds to our challenge question of \textit{obtaining a solution workflow for a DS and/or ML problem}. Practically, this transformation is most likely a combination of a domain expert \textit{building} and/or \textit{finding} workflows and workflow components.

That is, a domain expert could: a) find an existing workflow or components in a repository, b) rely on formal or informal mappings or recommendations to assemble a workflow, or c) build the workflow themselves either from a component library.

%These options are ordered here from immediate obtaining of the workflow to a time-consuming construction process.

Here we will summarise a few of the techniques available in the tools from Section~\ref{sec:tools} to assist a domain expert in obtaining or building a workflow.

\paragraph{Component Libraries}

Many of the graphical workflow tools (see Section~\ref{sec:tools_graphical}) use a library of components for the domain expert to select from when building their workflow. Plugins or extensions can extend this component palette with domain-specific components, allowing for easier selection of these components. For example, the Galaxy tool offers workflow components which directly obtain data from biology databases. This aids biologists to efficiently obtain up-to-date data directly into their workflows.

% \bentley{SHOW FIGURE HERE. }

\paragraph{Domain-specific Tutorials and Sample Workflows}

The documentation surrounding a workflow tool often provides numerous examples for using the tool and for solving real-world problems. For example, the Nipype tool discussed in Section~\ref{sec:tools_text} offers over 30 neuroscience-specific examples on its website. This allows users to get started quickly to solve their domain-specific problems.

\paragraph{Workflow Repositories}

Some of the workflow tools discussed in Section~\ref{sec:tools} have explicit \textit{repositories} for searching and obtaining workflows, or have multiple tutorial workflows for demonstrating the use of their components. These repositories allow for experts to select whole workflows and their component parts for use in solving their problem.

However, a brief glance at these repositories suggests usability issues. For example, one Galaxy workflow repository\footnote{\url{https://usegalaxy.eu/workflows/list_published}} contains hundreds of workflows, yet the only search options available cover \textit{free text}, \textit{user rating}, and \textit{keyword search}. This may make it quite difficult for a domain expert to find a suitable workflow unless techniques are used to further recover semantic information~\cite{Diaz2017}. Reproducibility issues may also hamper this search as computations may not be consistent between runs or user machines~\cite{Theaud2020}.

\paragraph{Automated Recommendations}

Recent work by Wen \textit{et al.} suggests that it may be possible to automatically determine similarity of workflows in repositories~\cite{Wen2020}. This could allow for enhanced discoverability of workflows.

\begin{figure}[tbh]
    \centering
    \includegraphics[width=0.25\textwidth]{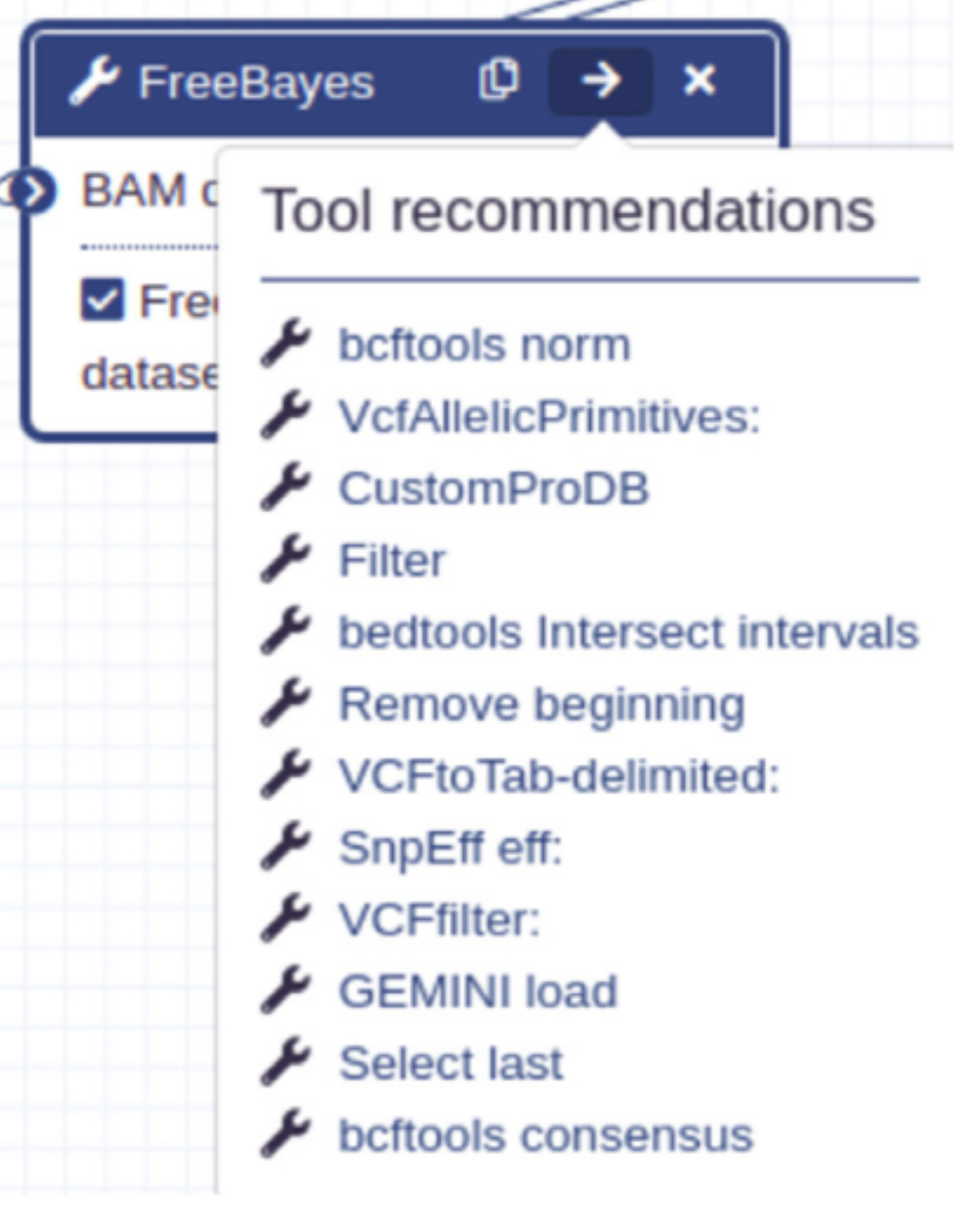}
    \caption{Automated recommendation of a component. Modified from~\cite{Kumar2021}.}
    \label{fig:galaxy_recommendation}
\end{figure}

Workflow tools can also recommend next components to be placed automatically. This allows for domain experts to be assisted by the tool to build their workflow. For example, the excellent article of Kumar \textit{et al.} presents a recommendation engine for the Galaxy framework~\cite{Kumar2021}. This engine integrates into Galaxy itself to provide component recommendations based on the existing components and the recent usage of that component in the data set. Figure~\ref{fig:galaxy_recommendation} shows the tool providing a list of recommended next components for the user to select from.

\paragraph{Automated Creation}

An interesting technique to create the whole workflow at once is to use machine learning techniques themselves to create workflows. This is the field of AutoML~\cite{He2021}, which uses accuracy metrics to create a machine learning pipeline for the domain expert's data. A partial or full workflow for the expert can also be provided based on their problem and/or their data~\cite{Gil2010,Gil2010a}. As a recent example of this AutoML approach, we point to Dunn \textit{et al.} who use a benchmarking set in materials science as the basis for creating material science-specific workflows~\cite{Dunn2020}.

\subsubsection{Region Transformation}

\begin{figure}[tbh]
    \centering
    \includegraphics[width=0.49\textwidth]{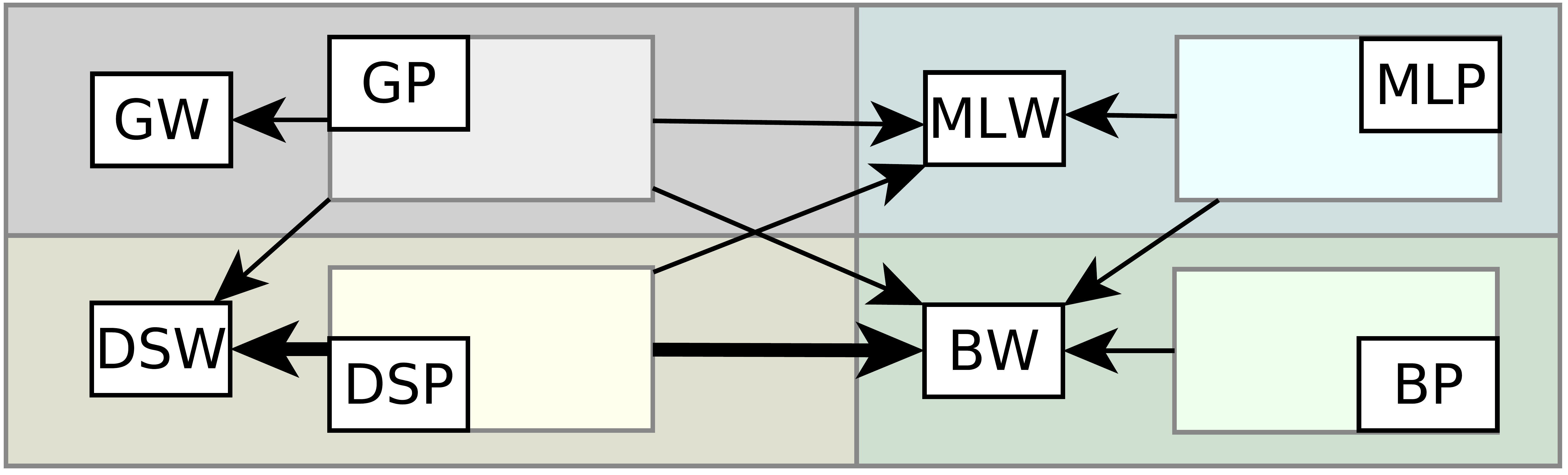}
    \caption{Transformations between the \textit{problem space} (inner boxes) and the \textit{solution workflow space} (outer boxes).}
    \label{fig:ps_to_ss_trans}
\end{figure}

Figure~\ref{fig:ps_to_ss_trans} diagrams some of the possible transformations between regions on the \textit{problem space} layer into the \textit{solution workflow layer}. That is, a \textit{domain-specific problem} (DSP) could be solved with a \textit{machine learning workflow} (MLW). As indicated in the figure with bolded arrows, the two preferable transformations are to take the original \textit{domain-specific problem} and directly create a \textit{domain-specific} or \textit{blended} workflow. This retains the domain-specific nature of the problem and adds the required machine learning components.

\subsubsection{Reversing the Transformation}

As mentioned above, it can be challenging for domain experts to search a workflow repository and find suitable workflows for their problem. One direction to address this issue is to automatically `mine' the \textit{workflow} itself and extract the \textit{problem} that workflow solves, or at least extract some tags and other semantic information~\cite{Diaz2017}.

\subsection{Solution Workflow Space to Implementation Space Transformations}

The transformations between the middle and bottom layers of the framework transform a \textit{workflow} in the \textit{solution workflow space} into some form of \textit{code} in the \textit{implementation space} (Section~\ref{sec:implementation_space}). This corresponds to our challenge question of \textit{producing an implementation from a workflow which is usable for a domain expert}.

The techniques examined here are: a) re-implement the workflow manually (not discussed below), b) code generation / Model-Driven Engineering (MDE) techniques, or c) workflow execution directly performed by the workflow tool.

\begin{figure}[tbh]
    \centering
    \includegraphics[width=0.49\textwidth]{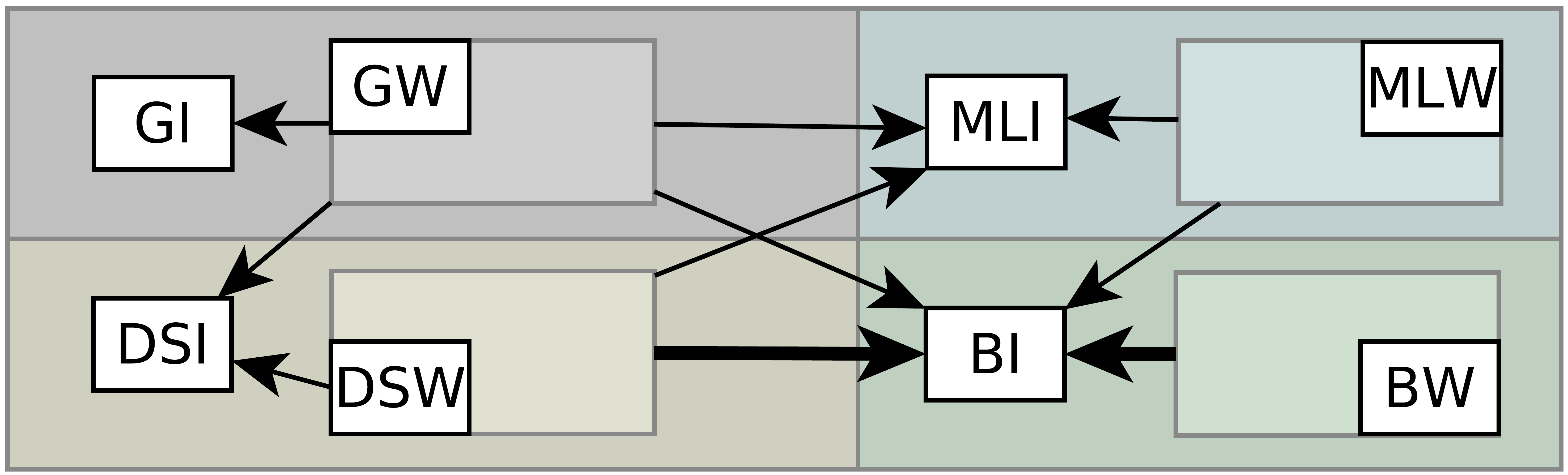}
    \caption{Transformations between the \textit{solution workflow space} (inner boxes) and the \textit{implementation space} (outer boxes.}
    \label{fig:ss_to_is_trans}
\end{figure}

\subsubsection{Model-Driven Engineering Techniques}

From a workflow defined in a DSL or as components, it can be a straightforward process to perform MDE techniques such as code generation~\cite{Banoo2020}. Due to the modular nature of the workflows, executable code could be generated for each component.

This code generation may also take place over a number of intermediate languages, such as using workflow middleware to handle concerns such as scalability (Section~\ref{sec:tools}).

\subsubsection{Direct Execution}

Another way of executing the workflow is to run it inside the workflow tool itself. This relieves the domain expert from running the final code themselves, though it may be more difficult to optimise if needed. Many of the tools in Section~\ref{sec:tools} execute the workflow in this way, either through the host language such as Python or within the tool itself such as Galaxy. This execution may also be local on the domain expert's machine or run on another machine.

\subsubsection{Reversing the Transformation}

An interesting direction of research is to consider taking legacy code and extracting the workflow from it as in our challenge question: \textit{extract a workflow from an existing implementation}. This could be a manual inspection or an automatic process. Once completed, the workflow could then become the source of truth and could be moved into a workflow repository for further dissemination and development~\cite{Carvalho2017}.

\section{Workflow Standards and Tools}
\label{sec:tools}

This section provides another contribution of this paper: an introductory (and therefore non-comprehensive) selection of workflow standards and tools available to experts from various domains. The intention is for readers to gain a quick understanding of the diversity and features of what is available today.

\paragraph{Limitations}

The primary limitation of this section is that no delve into the literature can reasonably cover the overwhelming number of workflow systems available. As a lower bound, Amstutz \textit{et al.} maintain an incomplete listing of more than 300 workflow systems~\cite{Amstutz2021}.

Instead, the selection in this report is to focus on \textit{open-source} \textit{scientific workflow tools} reported in the literature. That is, the standards and tools selected are not targeted towards business, application development, or those workflows which are built within a tool or platform\footnote{These workflows are commonly referred to as ``visual scripting''. For example, \url{https://unity.com/products/unity-visual-scripting}, \url{https://www.blackmagicdesign.com/ca/products/davinciresolve/fusion}, and \url{https://lensstudio.snapchat.com/guides/visual-scripting/}. Lens Studio is of particular interest with the integration of ML algorithms within the workflow.}. In this article, the field of biological sciences is very well-represented. This is due to the combination of big data and reproducibility concerns which motivated the creation of such tools~\cite{Wratten2021,Reiter2021}.

This report also aims to be a general introduction and does not touch on many of the ``ilities'' relevant for domain experts to use these tools. For example, Wratten \textit{et al.} evaluate twelve bioinformatic workflow managers using the categories of \textit{ease of use}, \textit{expressiveness}, \textit{portability}, \textit{ scalability}, \textit{learning resources}, and \textit{pipeline initiatives}~\cite{Wratten2021}. Admed \textit{et al} mention \textit{modularity} and \textit{reproducibility} amongst others~\cite{Ahmed2021}, while Kortelainen adds the important characteristics of \textit{licensing} and \textit{maturity}~\cite{Kortelainen2021}. Other factors may be connection to specialised tools or computing platforms such as the Hermes middleware platform for increased scalability~\cite{Kintsakis2017}.

\subsection{Workflow Formalisms and Standards}
\label{sec:tools_workflow_standards}

\textit{Workflows} can be represented in the simplest form as a connected and directed graph, where nodes in the graph are computations and the edges are dependencies of data or control. Extending beyond this representation are well-known formalisms which can also represent workflows.

For example, Petri Nets~\cite{Salimifard2001} can allow for formal verification of properties such as liveness for the system, or a Formalism Transformation Graph + Process Model (FTG+PM) can record the formalisms and transformations employed in the workflow~\cite{Challenger2020}. Another well-known workflow standard is the Business Process Model and Notation (BPMN)~\cite{Chinosi2012} to formalise both automatic and manual workflows within an organisation.

Bringing together both Petri Nets and BPMN is Yet Another Workflow Language (YAWL)\footnote{\url{https://yawlfoundation.github.io/}}. This workflow language from the 2000's takes Petri Nets as a starting point and adds extensions for commonly-seen workflow patterns~\cite{VanDerAalst2005}. The formal semantics of YAWL allow for verification of workflow properties such as \textit{soundness} (ensuring \textit{an option to complete}, \textit{proper completion}, and \textit{no dead transitions})~\cite{Wynn2009}.

More recently, a number of workflow standards have been developed in various sub-fields but none has yet established dominance over the others~\cite{Silva2021}. This may soon change with convergence on the Common Workflow Language (CWL)\footnote{\url{https://www.commonwl.org}}.

CWL originated in the bioinformatics community and offers a declarative workflow definition language (a DSL) that can be written in JSON or YAML to be executed by a workflow execution engine~\cite{Crusoe2021}. Of particular interest to this report is that DS attributes can be added to workflows and their steps as needed by users, allowing for a great deal of flexibility and discoverability for domain experts. The standard is becoming established throughout multiple domains and has a number of implementing tools, including upcoming support in the graphical Galaxy workflow tool discussed below.

\subsection{Workflow Tools and Management Systems}

Workflow tools and management systems can be related and considered as descendants from programming build systems~\cite{Leipzig2017} such as \textit{make}~\cite{Stallman2020} and SCons~\cite{Knight2005}. The objective is to record and manage the dependencies of each component, such that the computation can be correctly executed.

However, the wide variety of \textit{workflow management tools and systems} available today have an expanded set of concerns beyond compilation steps~\cite{Singh2019,Paul-Gilloteaux2021,Reiter2021}. This can include automatic versioning and provenance concerns, deployment to computational resources, and providing components for use in workflows. In particular, there is a strong usage of these systems on containerisation and package management to ensure that workflows can be re-executed in the same context that they were first developed in. The strong focus of these systems on the scientific requirement for reusability and reproducability is further discussed in Section~\ref{sec:discussion}. 

This section will touch upon some workflow systems found in use today and report some of the interesting features and considerations implemented. %\textit{Text-based} workflow systems will be discussed, divided into using a \textit{language module} or a \textit{domain-specific language} (DSL). Then \textit{graphical} workflow systems are presented.

\subsubsection{Text-Based}
\label{sec:tools_text}

Current text-based workflow systems seem to follow two approaches: either the system is implemented as a module/library for a general programming language such as Python, or the system ingests a standard \textit{markup language/DSL}.

% MLflow \url{https://www.mlflow.org}
%Ruffus \url{http://www.ruffus.org.uk/}
%\cite{Aly2018}

\paragraph{Language Module}

A common implementation strategy for workflow tools is to leverage the user's knowledge of a general programming language. Commonly, this is Python due to its widespread usage.

%\paragraph{Luigi}

For example, \textit{luigi} is a tool from Spotify\footnote{\url{https://github.com/spotify/luigi}} which allows a user to build up a dependency graph of \textit{Tasks} which interact with \textit{Target} files. These concepts are defined within Python code and the Luigi API offers access to common database/cloud tools. A web-based scheduler and visualiser is also available for monitoring long-running workflows.

Luigi was extended by Lampa \textit{et al.} into SciLuigi~\cite{Lampa2016} for scientific workflow requirements such as a separation of the workflow and the tasks, audit support, and support for high-performance computing. The authors then developed \textit{SciPipe}\footnote{\url{https://github.com/scipipe/scipipe}} in the Go programming language for enhanced type-safety and performance~\cite{Lampa2019}.

%\paragraph{atomate}

Workflow systems defined as language modules can also be tailored to particular domains, further reducing the amount of code a domain expert must write to use the workflows.

For example, the \textit{automate} tool\footnote{\url{https://atomate.org}} for computational materials science~\cite{Mathew2017} offers workflows to copy and customise based on specific analysis of materials, and an API to the analysis tools themselves. atomate uses the FireWorks workflow software which provides provence and reporting support for high-throughput computations~\cite{Jain2015}.

%\paragraph{nipype}

Another example of a domain-specific library for workflows is the \textit{nipype} Python software package\footnote{\url{https://nipype.readthedocs.io/en/latest/}} to define workflows in neuroimaging~\cite{Gorgolewski2011}. The intention here is to define components commonly used in neuroscience and have them as part of the same workflow. This allows domain experts who know Python to quickly build a workflow of neuroscience-specific tools.

Moving one level of abstraction higher, \textit{fMRIPrep}\footnote{\url{https://fmriprep.org/en/stable/}} is an automated workflow built on top of nipype~\cite{Esteban2019,Esteban2020}. fMRIPrep adapts to the input data automatically to performing the appropriate preprocessing steps for functional magnetic resonance imaging (fMRI), such as head motion correction and skull stripping. This assists in providing replicable results for neuroimaging studies both in terms of computation and by providing ``boilerplate'' natural language text for insertion into a research article's method section.

%\paragraph{MLOps and ZenML}

Replicable results are also important when considering the development of ML models. The emerging field of Machine Learning Ops (MLOps) tackles the automation, provenance, performance, and other aspects of ML in a workflow-based form~\cite{Ruf2021}. The \textit{ZenML} Python library\footnote{\url{https://zenml.io/}} provides a high-level API to machine learning tasks and tools, while offering workflow management features such as versioning, scheduling, and visualisation.

\paragraph{Markup or Domain-Specific Language}

The other workflow specification commonly seen in workflow management systems is to have a definition written in either a markup language (such as XML YAML) or a custom DSL for the workflow itself.

\textit{Compi}\footnote{\url{http://sing-group.org/compi/}} is a framework to not only build and run workflows, but also deploy the workflows as command-line applications or containerised as Docker containers~\cite{Lopez-Fernandez2021}. That is, once a domain expert has built a workflow, Compi packages the workflow and its dependencies can be easily shared to other domain experts to use as a command-line application. Compi uses the markup language XML to define the workflows as the creators L\'{o}pez-Fern\'{a}ndez \textit{et al.} argue that a DSL for defining workflows is ``less interoperable, being difficult to produce or consume from languages other than the one on which the DSL is based'', and that XML is ``easy to validate syntactically and semantically
through schemas''~\cite{Lopez-Fernandez2021}. A repository of Compi workflows is available through the Compi hub project\footnote{\url{https://sing-group.org/compihub/explore}} which aims to provide community exploration of the workflow, including automatic visualisation of the workflow tasks and links to sample input data~\cite{Nogueira-Rodriguez2020}.

\textit{Nextflow}\footnote{\url{https://www.nextflow.io/}} is a workflow management system ``designed specifically for bioinformaticians familiar with programming''~\cite{DiTommaso2017}. Workflows are designed in a Bash script-like DSL to manage data flow between different workflow components. The Nextflow tool itself has support for obtaining and setting up Docker containers to allow for greater reproducibility of workflows.

Nextflow also has an active ecosystem providing validated open-source pipelines. In particular, the nf-core effort\footnote{\url{https://nf-co.re/}} is a community-maintained effort to develop ``collaborative, peer-reviewed, best-practice analysis pipelines''\cite{Ewels2020}. Only one pipeline per data type/analysis is allowed, and all pipelines require quality checks such as a common structure, MIT licensing, continuous integration tests, linting, and appropriate documentation.

%Nextflow  See Pipeliner paper (Pipeliner: A Nextflow-Based
%Framework for the Definition
%of Sequencing Data Processing
%Pipelines)~\cite{Federico2019}

%\url{https://github.com/nextflow-io/awesome-nextflow}

%\url{https://biocontainers.pro/tools/abeona} Is this only Nextflow?

%Snakemake \url{https://snakemake.readthedocs.io/} See the repo at \url{https://github.com/snakemake-workflows}

\subsubsection{Graphical}
\label{sec:tools_graphical}

With the rise of ``low-code'' platforms, there are an increasing number of \textit{graphical} workflow systems available~\cite{Bock2021}. A prominent example of this is the domain of business applications, where providers such as outsystems\footnote{\url{https://www.outsystems.com/}} and Mendix\footnote{\url{https://www.mendix.com}} provide graphical interfaces to create applications which can involve ML.

%Dagster\url{https://github.com/dagster-io/dagster}

%Kelper\url{https://kepler-project.org/}

A workflow system straddling the domain-specific and business domains is the Konstanz Information Miner (KNIME)~\cite{Berthold2007}. From the University of Konstanz circa 2007, this framework originally focused on pharmaceutical applications\footnote{\url{https://www.knime.com/knime-open-source-story}} but has now scaled up for use within large-scale enterprises. KNIME is based on the Eclipse platform and offers a component library and canvas for drag-and-drop connection of nodes. KNIME also offers a repository for hundreds of components and workflows available for use with a selection of curated components available\footnote{\url{https://hub.knime.com/}}. Also relevant to this article is a feature within KNIME called the ``Workflow Coach''. From KNIME community usage statistics, this panel is able to recommend the next component to use in the workflow\footnote{See \url{https://www.knime.com/blog/the-wisdom-of-the-knime-crowd-the-knime-workflow-coach}}.

% as seen in Figure~\ref{fig:knime_workflow}
%\footnote{\url{https://www.knime.com/}}

% as seen on the left-hand side of Figure~\ref{fig:knime_workflow}
% \begin{figure*}[tbh]
%     \centering
%     \includegraphics[width=0.8\textwidth]{figs/tools/knime_workflow}
%     \caption{The KNIME editor showing a workflow for text processing.}
%     \label{fig:knime_workflow}
% \end{figure*}

%\cite{Chapman2021} details the definition of selecting patients for a disease such as COVID. Maps onto workflow.
%KNIME  \url{https://hub.knime.com/} \cite{Berthold2009}

The Workflow Instance Generation and Selection tool (WINGS)~\footnote{\url{https://www.wings-workflows.org/}} focuses on \textit{semantic workflows}, where each input and component has semantic information attached~\cite{Gil2010}. This information is represented in the form of triples which allows for domain-specific information to be used to select workflow components. WINGS can use this semantic information to select components, input datasets, and parameter values~\cite{Gil2010a}.

Focusing on more domain-specific workflow systems, we select three commonly-used graphical platforms: \textit{Node-RED}, \textit{Orange}, and \textit{Galaxy}.

\begin{figure*}[tbh]
    \centering
    \includegraphics[width=0.8\textwidth]{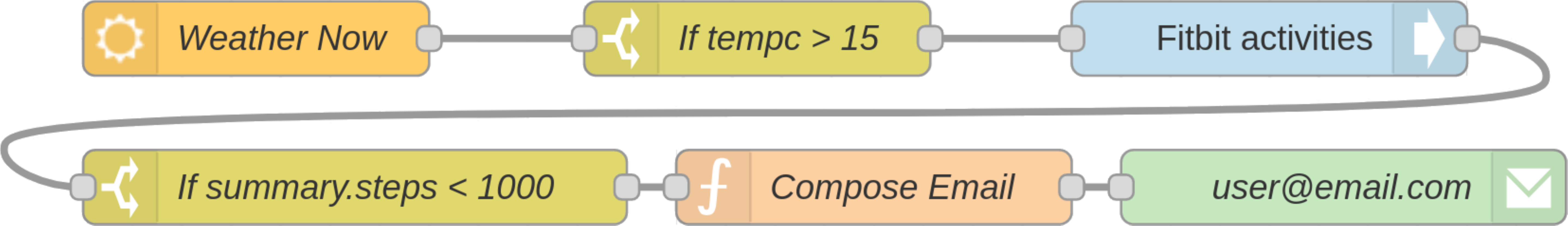}
    \caption{Example Node-RED flow to send an email when the weather is nice and a step counter is low. Figure modified from~\cite{NodeRedGuide}.}
    \label{fig:node_red_example}
\end{figure*}

\paragraph{Node-RED}

The \textit{Node-RED} tool is a web-based editor for creating Internet of Things (IoT) workflows~\cite{NodeRED}. Nodes provide built-in functionality or can be customised by adding Javascript code. Figure~\ref{fig:node_red_example} shows an example flow to remind a user to exercise. A node communicates with a weather service to obtain the temperature. If this value is above 15, then the flow communicates with an activity tracker. If this reports that a user's step count is low, then an email is sent. Flows can be deployed either to the user's local machine or many other embedded devices such as Raspberry PIs\footnote{\url{https://projects.raspberrypi.org/en/projects/getting-started-with-node-red/}}. Nodes and workflows can be shared in an online repository\footnote{\url{https://flows.nodered.org/}}, allowing users to enhance their workflows with new nodes\footnote{For example, ML nodes: \url{https://flows.nodered.org/node/node-red-contrib-machine-learning}.} and sub-workflows.

\begin{figure*}[tbh]
    \centering
    \includegraphics[width=0.5\textwidth]{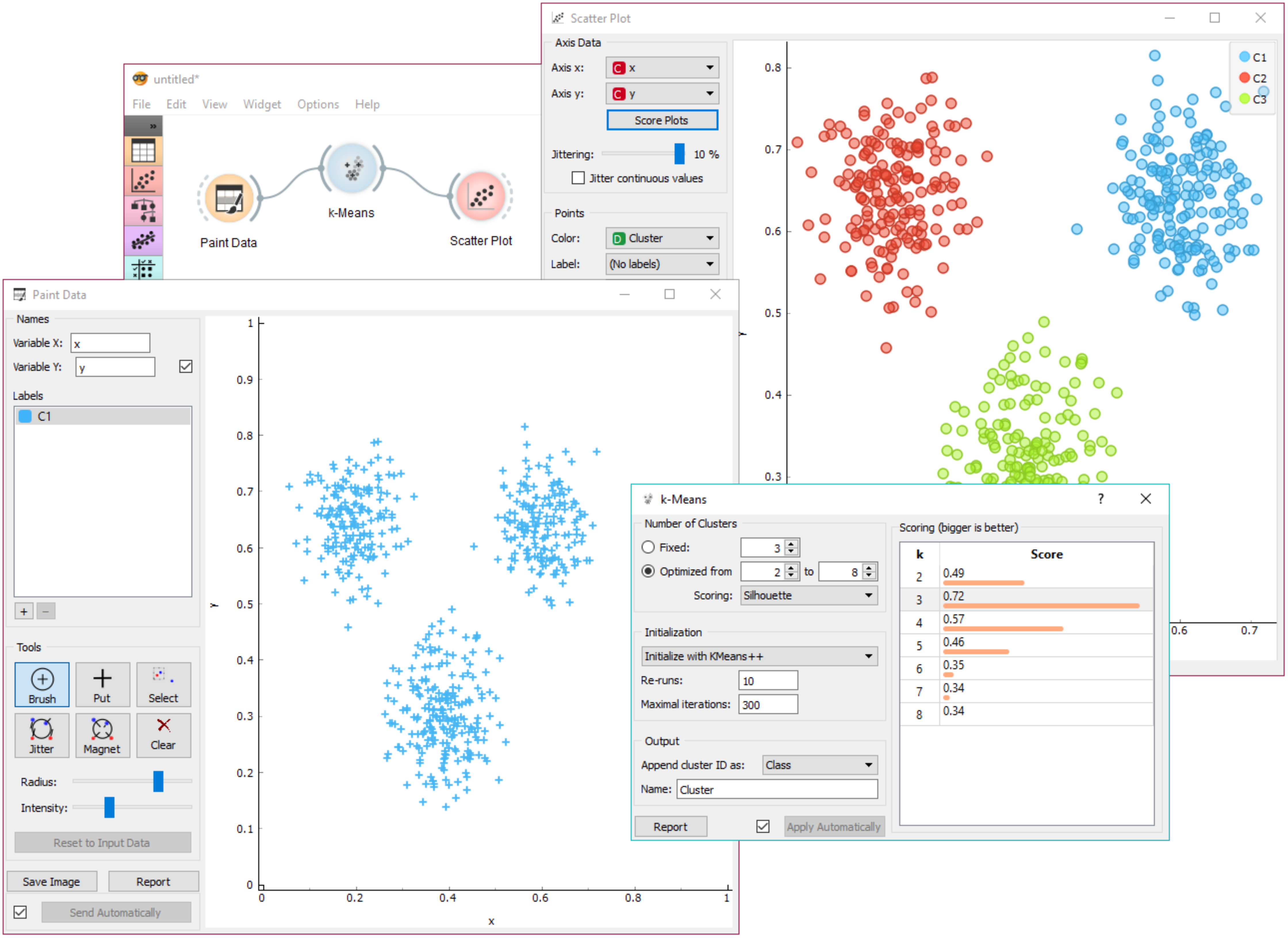}
    \caption{Example Orange workflow to visualise data, cluster it with K-Means, and then show the clustering~\cite{OrangeWebsite}.}
    \label{fig:orange_example}
\end{figure*}

\paragraph{Orange}
The \textit{Orange} tool offers visual scripting of data mining techniques including machine learning operators and visualisation capabilities~\cite{Demsar2005,Demsar2013}. Originating out of a bioinformatics research group, Orange focuses on providing an easy-to-learn data science tool. A canvas is provided for users to drag-and-drop nodes from a node library, where typed connections then aid users in assembling a workflow. Figure~\ref{fig:orange_example} shows an example workflow where data is visualised before clustering occurs with K-Means~\cite{OrangeWebsite}. Note the option pane for K-Means allowing a user to tune the parameters. The clustering is then visualised.

Orange has two features which improve the reuse of the tool and its workflows. First is the selection of workflows available on the Orange website\footnote{\url{https://orangedatamining.com/workflows/}}. This offers 20 sample workflows for performing common data science tasks such as performing principal component analysis. The second feature is the robust add-on support which provides new nodes for workflows (which are called ``widgets'' in Orange). For example, currently there are add-ons for bioinformatics, education, and explainable AI available from within Orange itself. Users can also create their own nodes to create DS workflows~\cite{Godec2019,Toplak2021}.

%Bonita \url{https://www.bonitasoft.com/bonita-platform}

\paragraph{Galaxy}
The \textit{Galaxy} project is a web-based entire computational workbench for developing biomedical workflows~\cite{Jalili2020}. It has spread to other fields as well with over 5000 publications citing Galaxy\footnote{See \url{https://galaxyproject.eu/citations} for publications focused on just one online instance of the workbench.}. The workbench heavily focuses on concerns such as reproducability, provenance of data and tools, and sharing of workflows and learnings.

\begin{figure*}[tbh]
    \centering
    \includegraphics[width=0.5\textwidth]{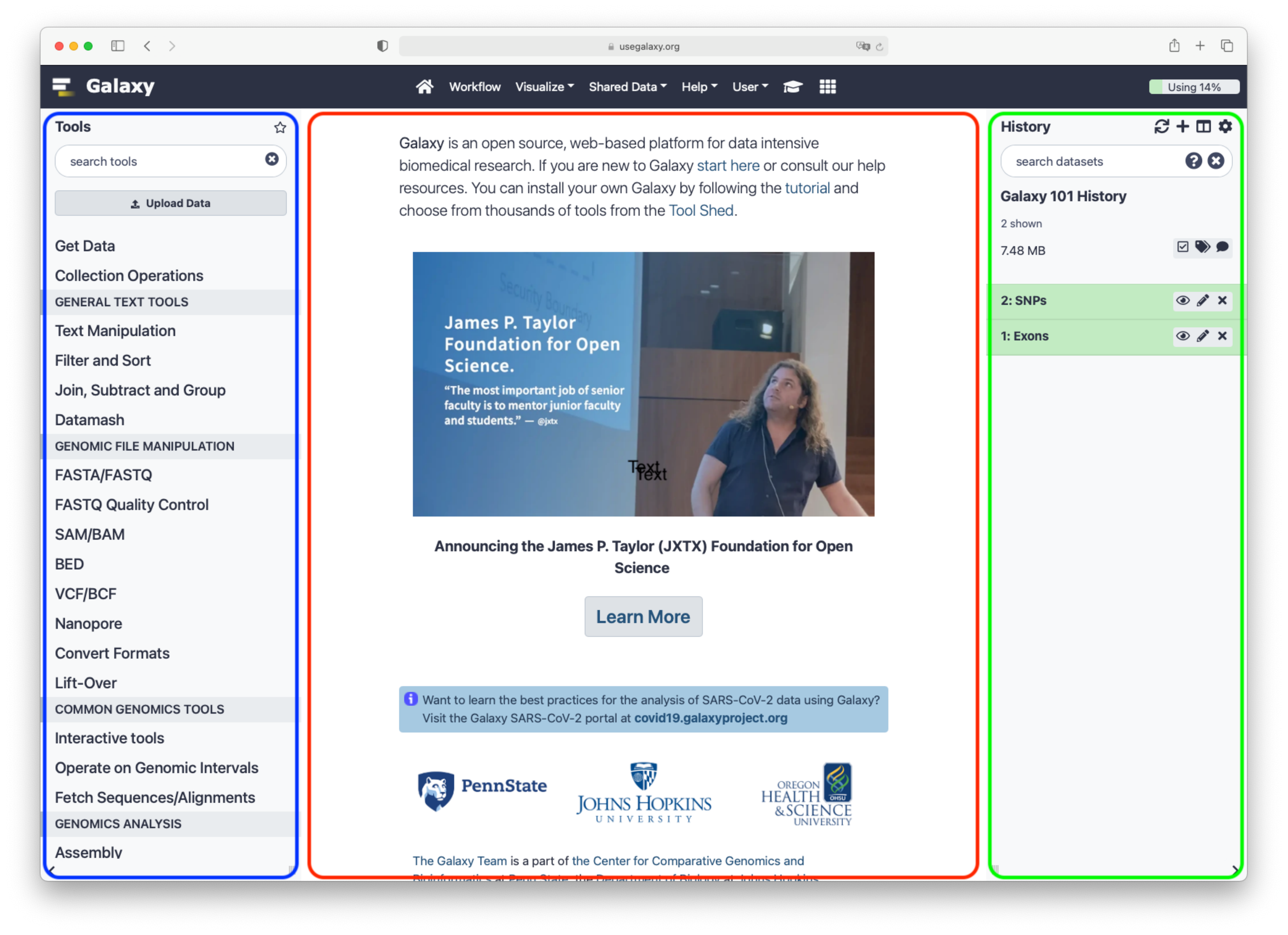}
    \caption{Example Galaxy interface with \textit{tool}, \textit{info}, and \textit{history} panes~\cite{Batut2018,Clements2021}.}
    \label{fig:galaxy_interface}
\end{figure*}

Figure~\ref{fig:galaxy_interface} shows the Galaxy interface from an on-line tutorial~\cite{Clements2021}. On the left is a pane for selecting tools which in this example focus on genomics. The centre pane is for displaying website, tool, or dataset information. The right pane details the history of the analysis. These histories correspond to inputs and outputs of the tools to be recorded and shared. These histories can also be visualised on a canvas. 

An interesting aspect of the Galaxy project is a focus on community and specialisation. For example, there is a main publicly-available Galaxy instance online\footnote{\url{https://usegalaxy.org/}}. However, to spread out computation costs and offer the possibility of specialising the datasets and tools available, Galaxy can run on a user's local or cloud machine. Furthermore, these other Galaxy instances can be customised into different \textit{workspaces}.

We highlight three recent DS specialisations of Galaxy which are publicly available online. Vandenbrouck \textit{et al.} developed a Galaxy instance for proteomics research~\cite{Vandenbrouck2019}, Tekman \textit{et al.} provide one for ``single cell omics''~\cite{Tekman2020}, and Gu \textit{et al.} offer a ML focus~\cite{Gu2021}. These specialisations each offer targeted DS tools, workflows, and computational resources, allowing domain experts to quickly develop workflows.

%\cite{Ramsey2020} Connecting Galaxy and Apollo

%\paragraph{Application-Specific}
%visual scripting
%\url{https://docs.godotengine.org/en/stable/getting_started/scripting/visual_script/getting_started.html} 
%\url{https://unity.com/products/unity-visual-scripting}
% Lens Studio \url{https://lensstudio.snapchat.com/}
% Da Vinci Resolve \url{https://www.blackmagicdesign.com/ca/products/davinciresolve/fusion}
% Dynamo BIM \url{https://dynamobim.org/}

\subsection{Relation to Framework}
\label{sec:tools_framework_relation}

The above tools focus on managing workflows for a user. In this section, we provide general comments relating these tools to the six questions specified in Section~\ref{sec:intro}. This is not intended to be a comprehensive analysis and comparison of the tools, but instead suggest which challenges are currently addressed, and which are not yet addressed by these tools.

\paragraph{Mapping a DS problem to a form suitable for ML}:
For the first question, none of the tools or their documentation address the issue of mapping a domain-specific problem to a ML one. This is understandable, as the workflow management systems focus on the workflow and implementation layers of our framework. However, further integration of problem specification and mapping into these tools may assist domain experts.

\paragraph{Providing a solution workflow for a DS and/or ML problem}:

This challenge is the core focus of these workflow management systems as they provide the domain expert with the formalisms and assistance to build up the workflow. However, it is clear that tools have different ways of assisting the user, as described in Section~\ref{sec:p_space_to_sw_space}. This includes assisted workflow composition, domain-specific examples, component libraries, and workflow repositories. Not yet fully addressed in all tools is automation techniques for constructing workflows such as recommending components.

\paragraph{Allowing the domain expert to experiment with appropriate ML components in a DS workflow}:

This challenge relates to the ease of which a domain expert can modify their workflow to include ML components. The workflow tools described here do not offer automated support for such a modification. However, tools such as Orange attempt to ease the experimentation process for domain experts through its use of typed component connectors, rich component library, and visualisation support.

\paragraph{Adding DS knowledge to improve ML performance}:

Only the WINGS tool was seen to improve the performance of ML techniques by utilising DS knowledge. This is possible due to semantic reasoning of domain knowledge which is used to select appropriate components and parameters. More ML-specific techniques such as feature extraction are possible but there does not seem to be integrated support for this challenge n the tools examined here.

\paragraph{Producing an implementation from a workflow which is well-suited for a domain expert}:

This challenge is very well addressed by the tools mentioned here. In particular, Galaxy offers powerful computational resources in a web-based platform. This allows bioinformatics experts to run their workflows on specialised platforms.

\paragraph{Extracting a workflow from an existing implementation (code or notebook}:

None of the tools offer support for extracting workflows from existing code. However, the light-weight nature of the Python module-based tools could be seen as an easy way to ``lift'' existing code into an explicit workflow.

\paragraph{Summary}

Table~\ref{tab:tool_support} offers a general analysis on whether the tools discussed in this section addressed the challenge questions. For each question, a summary of the techniques from Sections~\ref{sec:layers} and~\ref{sec:transformations} is presented. Symbols then provide an indication whether the challenge is \textit{not}, \textit{partially}, or \textit{more fully} addressed by the tools. The last column then highlights the best examples which address each challenge. 

\begin{table*}[tbh]
    \centering
    \begin{tabular}{l|p{5cm}|c|p{5cm}}
        \textbf{Question} & \textbf{Techniques} & \textbf{Addressed} & \textbf{Best Supporting Tool(s)} \\\hline

Mapping DS $\rightarrow$ ML & Expert mapping, ontologies & \hbempty & None\\

Problem $\rightarrow$ Workflow & Workflow composition, examples, repos, libraries & \hbfull & WINGS, Nextflow, Nipype, KNIME, Node-RED, Orange, Galaxy \\

Increasing workflow ML & Suggestions, experimentation & \hbrhalf & KNIME, Orange \\

Increasing workflow DS & Knowledge representation, feature extraction & \hbrhalf & WINGS\\

Workflow $\rightarrow$ implementation & Containerisation, deployment, run within tool & \hbfull & automate, nipype, Compi, KNIME, Node-RED, Orange, Galaxy\\

Implementation $\rightarrow$ workflow & Code mining, language integration & \hbrquarter & Nipype
         
    \end{tabular}
    \caption{Broad analysis of tool support for answering challenge questions.}
    \label{tab:tool_support}
\end{table*}

From Table~\ref{tab:tool_support} it is clear that there are a number of challenge questions which are not addressed in current workflow tools or their ecosystems. We also identify that while most tools offer support for constructing workflows from problems, this is not yet an automated process. Thus there are ample opportunities for improving the experience of domain experts to create solution workflows as discussed in Section~\ref{sec:discussion}.

% To provide a general and qualitative analysis, we score whether the question is \textit{not}, \textit{partially}, or \textit{very well} addressed by a tool either in the tool itself or through training materials such as a website. This offers both an overview of what questions are not addressed by this set of tools, and which tools may have best practices for that question.

\section{Case Studies}
\label{sec:case_studies}

This section examines the use of the tools from Section~\ref{sec:tools} in various domains. Cases studies selected from the scientific literature provide a sample of how experts in each domain are building domain-specific workflows utilising machine learning. This section thus serves to: a) ground our framework in the state-of-the-practice, and b) highlight research challenges and opportunities where the modelling community can assist domain experts.

As a caveat, these case studies have been selected in an \textit{ad-hoc} and non-systematic manner. Instead, the informal criteria was based on \textit{recent publication}, \textit{available artefacts}, and \textit{variety of domain}. The intention is to provide a flavour of the heterogeneity of the domains and the recent use of tools for discussion, not an extensive literature survey.

We focus on three sets of case studies in this section. The first set is where the case study has an \textit{implicit} workflow. That is, there is no explicit workflow graph in one of the standards or tools reported in Section~\ref{sec:tools}, and the workflow is expressed in code. The second set of case studies contains \textit{explicit} workflow artefacts, where the workflow is explicitly defined using a workflow standard/tool. The third set is one case study where the high-level workflow itself is implicit, but it relies on a sub-workflow defined using a workflow framework.

% As discussed in Section~\ref{sec:discussion} we observe great potential on making these workflows more explicit, and improving the domain expert's ability to select domain-specific or machine learning components for their workflows.

%For each of the case studies below, the ``path'' through the framework from Section~\ref{sec:overview} is discussed. The transformations employed will be discussed, along with the benefits and drawbacks of each route.

\subsection{Case Study Overview}

Table~\ref{tab:case_studies} lists the case studies examined in this report and discussed throughout this section.  Each case study is provided an identifier based on the primary tool used and a short description. The second column in Table~\ref{tab:case_studies} denotes whether the case study has an \textit{implicit}, \textit{explicit}, or \textit{hybrid} workflow, and the third column lists whether the workflow is created \textit{textually} or \textit{graphically}. The last column reports the regions through our framework that the case study has artefacts in. That is, what ``path'' the authors took through our framework from Section~\ref{sec:overview}. This can be \textit{ML}, \textit{DS}, or \textit{B} for \textit{blended}. The Kaggle case also shows that a blended workflow can have a strong lean towards one dimension. In this case the lean is towards the ML dimension, represented by \textit{BML}.

%The last column reports whether the case study is likely to produce a workflow and/or implementation which focuses on the \textit{domain-specific} or \textit{machine learning} dimension, or \textit{either}. That is, is the workflow provided mostly dealing with DS concepts, ML concepts, or both.

\begin{table*}[tbh]
    \centering
    \begin{tabular}{l|p{8cm}|l|l|l}
    \textbf{Label} & \textbf{Description}  & \textbf{Repres.} & \textbf{Workflow} & \textbf{Regions}\\
    & & &  \textbf{Expression} & \\\hline
    
    CS\_PyTorch & Detecting peaks in \textit{metabolomic} data & Textual & Implicit & B-B-B\\
    
    CS\_MATLAB & Classifying rock origin based on molecular structure (\textit{geochemistry}) & Textual & Implicit  & B-B-B\\
     
     CS\_Kaggle & Crowd-sourcing ML solution to \textit{marine biology} challenge & Textual & Implicit & ML-BML-BML\\
     
    %CS\_Nextflow & Reproducible dMRI tractography workflow (\textit{neuroscience})  & Textual & Explicit  & Either\\
    
    CS\_nipype & Detecting depression from MRI images (\textit{neuroscience}) & Textual & Hybrid & B-DS-DS\\

    CS\_Orange & \textit{Data mining analysis} for smart school Internet traffic. & Graphical & Explicit  & ML-ML-ML\\
    
    CS\_Galaxy & Classification of urothelial cancer (\textit{bioinformatics}) & Graphical & Explicit & DS-DS-DS\\\hline
    
    %CS\_Ideal & A
    
    \end{tabular}
    \caption{Summary of the case studies.}
    \label{tab:case_studies}
\end{table*}

% \begin{figure*}[tbh]
%     \centering
%     \includegraphics[width=0.6\textwidth]{figs/case_study_flow}
%     \caption{The transformations within the case studies.}
%     \label{fig:case_study_flow}
% \end{figure*}

For each case study, the domain-specific problem is described. Then, the regions and transformations from our framework (Section~\ref{sec:overview}) which are relevant are presented.% Finally, each case study concludes with our remarks.

%and pointers to other interesting related works.

\subsection{Implicit Workflow Case Studies}

The first four case studies presented focus on \textit{implicit} workflows where there is not a workflow explicitly defined in one of the standards or tools from Section~\ref{sec:tools}. These case studies presented therefore cover cases where the domain expert directly writes an implementation of their problem, skipping the workflow layer of our framework (Section~\ref{sec:overview}). A discussion of these implicit versus explicit workflows is found in Section~\ref{sec:discussion}.

\subsubsection{CS\_PyTorch}

The first case study we examine represents the situation where a domain expert encodes their problem directly upon a ML library such as PyTorch or sklearn. 

\paragraph{As a Suggested Practice}

Directly writing code on an ML library is at a low-level of abstraction requiring a great deal of ML knowledge. However, we have found two recent publications from 2020 and 2021 where this approach is suggested.

For chemistry students, Lafluente \textit{et al.} present an introductory workshop focusing on utilising Python and visualisation/ML libraries~\cite{Lafuente2021}. The example Jupyter notebooks~\footnote{\url{https://github.com/ML4chemArg/Intro-to-Machine-Learning-in-Chemistry}} lead students through an introduction to Python, basic statistics, exploratory data analysis, classification, and regression.

In the field of materials science, Wang \textit{et al.} suggest that utilising Python code and the PyTorch library is considered `best practice'~\cite{Wang2020}. The example Jupyter notebooks\footnote{\url{https://github.com/anthony-wang/BestPractices}} walk a domain expert through an example application to highlight ML techniques and considerations at a very granular level of detail. For example, the reader is taken through constructing the layers of a neural network in PyTorch, along with calling the prediction/back-propagation functions. While this work is very comprehensive and suggests many useful and concrete suggestions for utilising ML in materials science, we (kindly) suggest that this is the wrong level of abstraction for a domain expert to utilise ML at. While the libraries themselves already abstract the low-level details, raising the level of abstraction further may be more appropriate for non-programmers.

Many factors may require utilising ML techniques at this low level of abstraction may be required for functional properties. For example, obtaining high degrees of parallelism is cited by Zhou \textit{et al.} as one reason for building a material science library upon PyTorch~\cite{Zhou2020}. The PYSEQM library\footnote{\url{https://github.com/lanl/PYSEQM}} implements functions applicable for semi-empirical quantum mechanics models, offering a high-level and domain-specific interface which is able to compute over the wide variety of GPUS which PyTorch supports, offering a speedup over other tools.

\paragraph{Case Study}

As the publications from Lafluente \textit{et al.} and Wang \textit{et al.} are targeted toward chemical science researchers just beginning to utilise ML, we also present here a case study of a chemical analysis tool \textit{peakonly} utilising ML.

\begin{figure*}[tbh]
    \centering
    \includegraphics[width=0.5\textwidth]{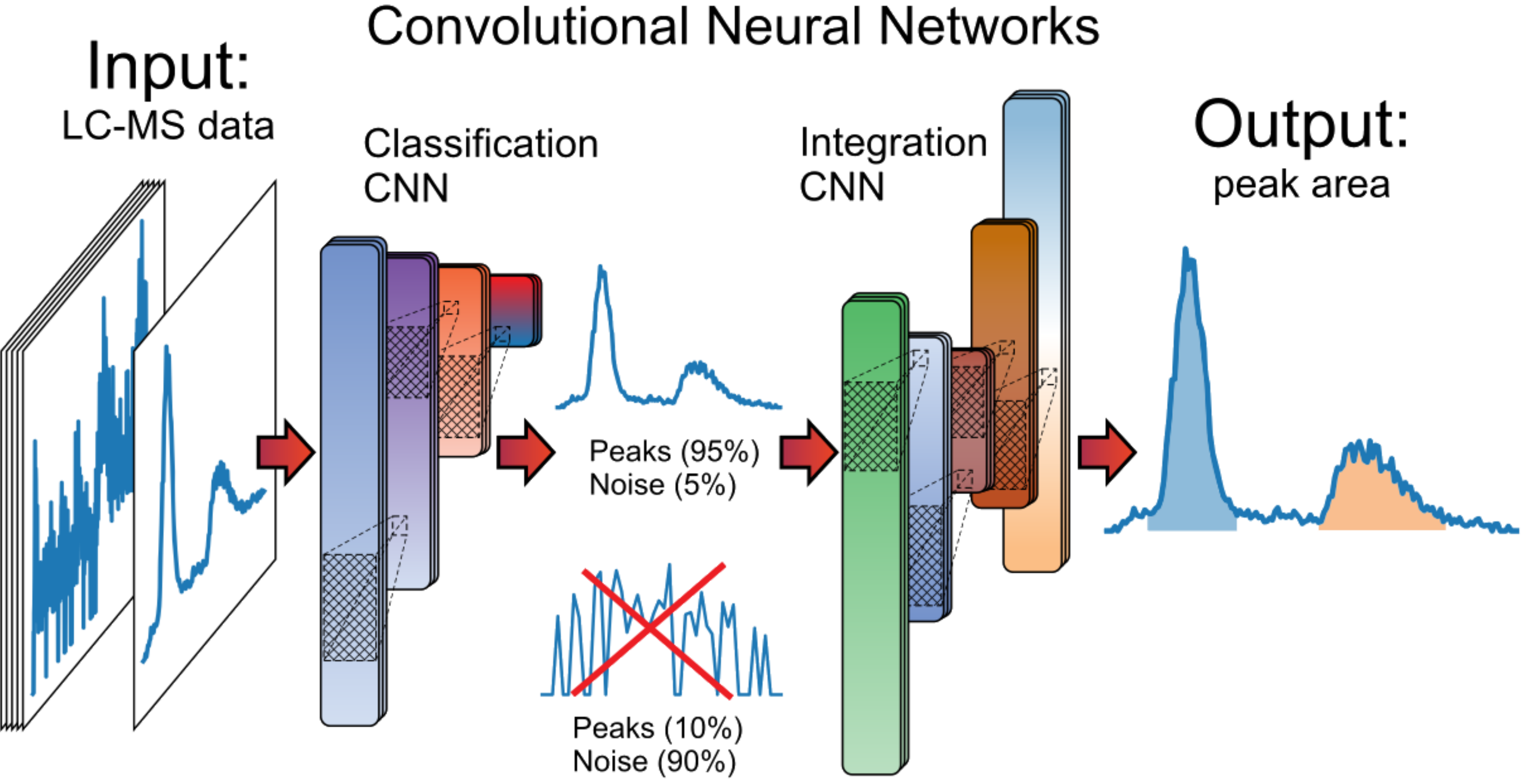}
    \caption{Visual abstract from Melnikov \textit{et al.} demonstrating classification and integration of peaks~\cite{Melnikov2019}.}
    \label{fig:melnikov_abstract}
\end{figure*}

Melnikov \textit{et al.} present an application of deep learning to classifying and integrating peaks in raw liquid chromatography-mass spectrometry (LC-MS) data~\cite{Melnikov2019}. The problem studied in the work is how to detect regions of interest (peaks) which occur in the noisy LC-MS data. Figure~\ref{fig:melnikov_abstract} shows how the data is first classified by a convolutional neural network (CNN) as a) noise, b) one or more peaks, or c) needs manual classification\footnote{Figure reprinted (adapted) with permission from~\cite{Melnikov2019}. Copyright 2020 American Chemical Society.}. A second CNN then provides the integration boundaries. The results from the CNN are validated against another tool. The authors provide a graphical tool \textit{peakonly} to perform these actions and visualise the output.

\paragraph{Problem Layer}

The DS problem specified in the case study is to detect and integrate the peaks in the data. The authors have transformed this into a \textit{blended} problem by utilising the \textit{expert mapping} given by prior work, where deep learning is used to perform peak detection and noise filtering.

\paragraph{Solution Workflow Layer}

Figure~\ref{fig:melnikov_abstract} displays a high-level view of the paper's approach to peak classification and integration. This forms the basis of the workflow components which we extract in Table~\ref{tab:cs_pytorch_workflow}. Note that these steps are performed by a user interacting with the \textit{peakonly} graphical interface, though a runner script is available.

\begin{table*}[tbp]
    \centering
    \begin{tabular}{l|p{2cm}|p{9cm}|c}
        \textbf{High-level Step} & \textbf{Step} & \textbf{Description} & \textbf{Classification} \\\hline
        
        Obtain Data & Load file & Load .mzML input data into tool & DS\\
        & ROI detection & Run algorithm to detect ROIs & DS\\\hline
        Calculate & Classify & Run classification with CNN & ML\\
        & Integrate & Run integration with CNN & ML\\\hline
        Analysis & Plot & Plot integrated ROIs & DS\\
        & Write out & Write output CSV file & General\\
        
    \end{tabular}
    \caption{Implicit workflow components in \textit{peakonly} tool (Melnikov \textit{et al.})~\cite{Melnikov2019}}
    \label{tab:cs_pytorch_workflow}
\end{table*}

Table~\ref{tab:cs_pytorch_workflow} is our attempt at extracting the steps in the workflow from the publication and tool of Melnikov \textit{et al}. The first column is the high-level step which summarises the individual steps in the second column. The \textit{description} column details what the step performs, with the quoted text copied from the code to explain the DS steps. The last column in Table~\ref{tab:cs_matlab_workflow} is our rough classification of whether the step is \textit{general} (generic programming code), \textit{domain-specific} (DS), \textit{machine learning} (ML), or \textit{blended}.

First, the DS file format is loaded into the tool. Then an algorithm runs to detect the regions of interest (ROIs). These two steps are highly domain-specific. The two CNNs are executed to first classify the ROIs and then to integrate the ones with detected peaks. We classify these as ML-intensive steps. Finally, the user is presented with the ROIs in a list where they are able to visually inspect the plot. The data can then be exported to a CSV file.

From our classification of these steps, it is fair to say that this is a \textit{blended} implicit workflow. There is a balance of DS and ML components.

\paragraph{Implementation Layer}

The implementation for the \textit{peakonly} tool is in the Python language, with usage of common data science/ML libraries (matplotlib, numpy, pandas, scipy, PyTorch) as well as the DS library pymzML for mass spectrometry data. This code can be classified as \textit{blended} due to the extensive mixed use of these libraries within the tool.

\paragraph{Remarks}

The \textit{peakonly} tool is an excellent example of domain experts utilising ML to solve a DS problem and providing an easy-to-use graphical interface to it. This lowers the barriers to entry for other domain experts to utilise ML on their similar problems.
\subsubsection{CS\_MATLAB}

For the second case study which focuses on using the common MATLAB tool which combines data processing and ML techniques, we have selected a publication %from the geochemistry domain. 
%The first publication is 
from Hasterok \textit{et al.} in the field of geochemistry~\cite{Hasterok2019}. The studied problem is to use the chemical composition of metamorphic rocks to classify whether the origins of the rocks were sedimentary or igneous. The resulting MATLAB code is available on GitHub\footnote{\url{https://github.com/dhasterok/global_geochemistry/tree/master/protolith/}}.

%The second publication is from Song \textit{et al.} in the field of chemical engineering~\cite{Song2020}. The issue they address is to predict the amount of C$O_2$ which can dissolve in an \textit{ionic liquid}, which is a salt which is liquid at room temperature. 

 %For the publication of Song \textit{et al.}, the code is available as supplementary data to the article\footnote{\url{https://www.sciencedirect.com/science/article/pii/S0009250920302840}}.

\paragraph{Problem Layer}

On the problem layer, there is a clear DS problem of classifying rocks on their chemical structures.

%, and b) predicting C$O_2$ solubility based on the molecular structure of a ionic liquid.

However, the authors venture further into the ML domain to determine which available ML classifiers are best suited for their problem. %In Hasterok \textit{et al.}, the authors
They describe investigation of principal component analysis (PCA) to filter the data before classification, as well as comparing K-nearest neighbour, decision trees, ensemble trees, and testing various parameters within these classifiers.

%In Song \textit{et al.}, the problem is extended to determine whether an artificial neural network (ANN) or a support vector machine (SVM) is more accurate for classification.

A robust knowledge about applying ML to their DS problem is shown. Thus we can classify this paper as addressing a \textit{blended} problem. The method of transforming the DS problem into a blended problem is not detailed, but based on the extensive related work cited we surmise that the authors gained this knowledge by reading past work which is a form of \textit{expert mapping}.

\paragraph{Solution Workflow Layer}

As mentioned, the solution provided by the authors does not contain an explicit workflow. Instead, the MATLAB code forms an implicit workflow operating on the input data and resulting in a classification or prediction. The authors provide their code in separate MATLAB scripts with comments, allowing us to reconstruct the predictor workflow and divide the scripts into DS and ML. Note that this table contains workflows for both the training and prediction processes.

\begin{table*}[tbp]
    \centering
    \begin{tabular}{l|p{2cm}|p{9cm}|c}
        \textbf{High-level Step} & \textbf{Step} & \textbf{Description} & \textbf{Classification} \\\hline
        
        Obtain Data & Load learning data & Data loaded into MATLAB & General\\
        & Treat learning data & \texttt{prep\_for\_cluster.m} selects data for training and testing. Can select to use meta-igneous/meta-sedimentary rocks or not & Blended\\\hline
        Training & Training & Using code generated from MATLAB's Classification Learner app & ML\\
   Analysis     & Write out & Write out model files & General\\
  & Plot & Plot training performance & General\\\hline
        
        Prediction & Convert and read predict data & Converting XLS data to CSV, and reading & General\\
        & FEFIX & ``Convert all Fe to FeO and calculate Fe2+/Fe\_total ratio'' & DS\\
        & CAT2OX & ``Convert cations to oxide data when missing'' & DS\\
        & OXIDE\_NORM& ``Computes the oxide norm for a set of given oxides'' & DS\\
        & Prediction & Use MATLAB \texttt{predict} function with the classifier and input data & ML\\\hline
        Analysis & Write out & Write to CSV file & General\\
        & Analyse & Plot classification performance & General\\
        
    \end{tabular}
    \caption{Implicit workflow components in MATLAB scripts for predicting protoliths (Hasterok \textit{et al.})~\cite{Hasterok2019}}
    \label{tab:cs_matlab_workflow}
\end{table*}

From this overview of the workflow, we can conclude that there is a mix of both DS and ML components. Thus this is a \textit{blended} workflow.

\paragraph{Implementation Layer}

Following the classification of the implicit workflow as \textit{blended}, it is clear that the implementation is also \textit{blended}.

This code was (presumably) hand-written by the authors. The exception is the code in the \textit{training} step\footnote{\texttt{train\_RUSBoost\_Classifier\_30l\_1000s\_20190222.m}.}. This code seems to have been generated by the MATLAB Classification Learner app\footnote{\url{https://www.mathworks.com/help/stats/classificationlearner-app.html}}.

\paragraph{Remarks}

From reading the publication of Hasterok \textit{et al}, it is clear that the authors have obtained a great deal of ML knowledge in addition to their domain expertise. They discuss pre-processing the data through principal component analysis (PCA) and perform a comparison between multiple ML classification techniques. This knowledge of both domains is reflected by the classifications given by our framework, where all of the problem, solution workflow, and implementation are blended.

The authors also leverage the ML functions built-into MATLAB for performing the training and classification, including the use of a MATLAB app to generate the appropriate training code.

The code provided contains an implicit workflow defined in MATLAB code. However, the authors have taken the time to modularise this code into various functions. This improves the usability and reproducibility of the code amongst other researchers.

\subsubsection{CS\_Kaggle}
\label{sec:cs_kaggle}

\textit{Kaggle}\footnote{\url{https://kaggle.com/}} is an online data science platform allowing data scientists to share models and code. Kaggle is well-known for its ``challenges'', where an organisation posts their data and an evaluation metric, and asks the Kaggle community to come up with a solution to that metric. For example, the ``NFL Big Data Bowl'' challenge asked for a prediction of how far one team will advance on the field during one play~\cite{Gordeev2020}. The data provided contained position, speed, and rotation information for each player on the field, weather information such as temperature, humidity, and wind velocity, and other data points. Once the competition has ended, the organisation sometimes contact the winners to explain their solution.

There are challenges in utilising Kaggle as a crowdsourcing tool to provide ML solutions for DS problems~\cite{Bumann2021}. Briefly, the benefits of this approach is that domain experts can directly interface with ML experts on Kaggle-hosted forums to share best practices and domain-specific information. The hope is that this will lead to knowledge transfer and provide high-quality insights for the domain experts, and offer money, prestige, and valuable skill training for the ML experts~\cite{Tauchert2020}.

There are a number of drawbacks, however. The domain experts have to spend effort to set up the contest by providing prize money and accurate problem data. Also, the solution provided by the ML experts may not be immediately applicable to the DS problem and lessons learned must be transferred back to the domain experts.

For example, in the Killer Shrimp Challenge a trivial solution was found by participants. The data was organised such that all data points with an index value greater than or equal to 2917769 had the presence of the killer shrimp. Thus the solutions of participants could simply test the index of each data point to obtain perfect accuracy on the predictions.

As an example of the effort required to transfer the ML lessons learned back to the domain experts, we point to the excellent article of Sutton \textit{et al.}~\cite{Sutton2019}. This article examines the top three solutions of a materials science challenge to determine the impact of the representation versus the learning method on the final accuracy. They describe how the first-place solution was a novel representation for material property ML.

\paragraph{Problem Layer}

Returning to the Killer Shrimp Challenge, the underlying problem is to detect the presence of the species \textit{Dikerogammarus villosus} (also known as the ``Killer Shrimp'') which causes environmental damage and is invasive in Europe. The specific Kaggle challenge is to take data points on water salinity, temperature, depth, wave exposure, and the presence of sand, and predict whether the Killer Shrimp will be present or not~\footnote{\url{https://www.kaggle.com/c/killer-shrimp-invasion}}. We refer readers to the article of Bumann \textit{et al.}~\cite{Bumann2021} for a full description of the challenge set-up and interactions between the domain experts and challenge participants.

It is interesting to note that the DS problem of predicting the presence of Killer Shrimp was effectively turned into a problem of predicting 1 or 0 in a particular column of a spreadsheet. Thus we classify this as a \textit{DS} problem being mapped into a \textit{ML} problem due to the lack of DS concepts in the problem statement.

\paragraph{Solution Workflow Layer}

In this report, we will attempt to extract the (implicit) workflow from the publically-available second-place solution~\cite{KaggleNotebook}. The solution is available as a Jupyter notebook which aids in the reconstruction of the workflow.

Table~\ref{tab:cs_kaggle_workflow} displays our extraction of the workflow in the solution. Note that similar to other workflows, we define the loading and saving of CSV files as rather general steps. Most of the remaining steps are solely ML-focused. However, the author of the notebook has also added two DS columns. The first added feature is \textit{water density} which is calculated from the temperature, salinity, and depth values. The second column is a classification of the wave exposure value to record whether the point is extremely or very exposed to waves.

\begin{table*}[tbp]
    \centering
    \begin{tabular}{l|p{2cm}|p{9cm}|c}
        \textbf{High-level Step} & \textbf{Step} & \textbf{Description} & \textbf{Classification} \\\hline
        
        Obtain Data & Load files & Load .csv files & General\\
        & Clean data & Fill in missing data with sklearn.impute.IterativeImputer & ML\\\hline
        Feature Adding & Add Density & Calculate ocean density based on other columns & DS\\
        & Classify Exposure & Classify wave exposure into categories & DS\\
        & Add Temperature Feature & Add column based on trained polynomial of temperature & ML\\
        & Add Outlier Feature & Add column to determine if data point is outlier & ML\\\hline
        Train & Train Classifier & Train a classifier & ML\\\hline
        Predict & Prediction & Predict the presence & ML\\\hline
        Analysis & Write out & Write output CSV file & General\\
        & Plot Feature & Plot the importance of features & ML\\

    \end{tabular}
    \caption{Workflow components in a Jupyter notebook for the Killer Shrimp Challenge~\cite{KaggleNotebook}.}
    \label{tab:cs_kaggle_workflow}
\end{table*}

In summary, this workflow is mostly comprised of ML components. However, due to the presence of these added DS features, one could say it is a \textit{blended workflow} with a heavy focus towards the ML side.

\paragraph{Implementation Layer}

The Jupyter notebook solution contains Python code and imports the expected data science/ML libraries (numpy, pandas, matplotlib, sklearn, xgboost). As the workflow is \textit{blended} (though tilted towards ML), the implementation can be said to be \textit{blended} as well.

\paragraph{Remarks}

It was surprising to see features added to the data which were DS. Our expectation was that this sort of DS knowledge would not be as present in Kaggle solutions, based on the expertise focuses of the challenge organisers and the ML experts. For example, we wish to highlight this quote from the analysis of Gordeev and Singer on their winning entry for the football challenge~\cite{Gordeev2020}:

\begin{quote}
    Don’t worry about having domain knowledge to attempt a specific problem. The main thing we learned in this competition is that you don’t necessarily need domain knowledge or industry [sic] to successfully tackle the data science challenge. Sometimes it even can be an advantage, as you go in blindly without many prior assumptions that might wrongly steer your exploratory analyses.
\end{quote}

Thus, providing a transformation from the DS problem to a ML representation may have to be balanced between prior DS knowledge and ML analyses.

\subsection{Hybrid and Explicit Workflow Case Studies}

The remaining case studies presented here are those which explicitly represent the workflow (or a sub-workflow) in one of the standards or tools from Section~\ref{sec:tools}. As discussed in Section~\ref{sec:discussion}, an explicit workflow aids with reproducibility, modularisation, collaboration, re-use, etc.

We also note that the explicit workflows tend to have a strong focus on enabling plugins or extensions for domains. This means that users are able to customise the workflows for their domains easily.

\subsubsection{CS\_nipype}
\label{sec:cs_nipype}

% Liu X, Bienkowska JR, Zhong W (2020) GeneTEFlow: A Nextflow-based pipeline for analysing gene and transposable elements expression from RNA-Seq data. PLoS ONE 15(8): e0232994. \url{https://doi.org/10.1371/journal.pone.0232994}
% \url{https://github.com/zhongw2/GeneTEFlow}

Practitioners may use workflow management systems such as \textit{nipype} to develop reproducible workflows focusing on particular domain processing tasks. For example, Celestine \textit{et al.} present a Python module\footnote{\url{https://github.com/sammba-mri/sammba-mri}} for performing pre-processing workflows such as DS file conversion and skull stripping for small mammal MRI brain data~\cite{Celestine2020}. 

These pre-processing tasks are essential for transforming the data such that it can be treated with ML. In this case study, we analyse a workflow which uses fMRIPrep as a sub-workflow to process MRI data before it is used to predict whether the person in question has depression~\cite{Mousavian2021}. 
As mentioned in Section~\ref{sec:tools_text}, this fMRIPrep automated workflow is built on top of nipype to perform pre-processing of fMRI data~\cite{Esteban2019,Esteban2020}. As such, the fMRIPrep tool itself is an \textit{explicitly} defined workflow. However, the authors of Mousavian \textit{et al.} have defined an \textit{implicit} workflow in Python to orchestrate the usage of the fMRIPrep tool. Thus this is an interesting \textit{hybrid} workflow case study which utilises an explicit tool workflow.

\begin{table*}[tbp]
    \centering
    \begin{tabular}{l|p{3cm}|p{8cm}|c}
        \textbf{High-level Step} & \textbf{Step} & \textbf{Description} & \textbf{Classification} \\\hline
        
        Pre-Process Data & Load data & Load data into Python & General\\
        & Convert format & Convert into BIDS format & DS\\
        & fMRIPrep & fMRIPrep workflow for alignment/timing correction & DS\\
        & FSL FEAT & Smoothing/filtering steps & DS\\\hline
   Defining Features     & Data Preparation & ``Background removement, normalise, resize, \& cube definition'' & DS\\
  & Prep. Initial Features & Correlate cubes & DS\\
  & Feature Selection & Rank correlations with stat. tests & ML\\\hline
  Classification & Classification & Run classifiers on data & ML\\
        
    \end{tabular}
    \caption{Workflow components for predicting depression from MRI images (Mousavian \textit{et al.})~\cite{Mousavian2021}}
    \label{tab:cs_nipype}
\end{table*}

\paragraph{Problem Layer}

The specified problem of Mousavian \textit{et al.} is to classify MRI images on whether the subject has Major Depression Disorder (MDD) or not. There are three major challenges addressed in the article. The first is to investigate different correlation measures of the \textit{voxels} (essentially three-dimensional pixels) of the MRI data. These correlation measures relate different areas of the brain together, and are used as features for the ML classification task. The second challenge is to handle imbalanced data sets where many subjects within the set do not have depression which can cause issues with classification. The third major challenge is to determine which of 14 ML classifiers performs best on the dataset.

From the problem and these specified challenges, it is clear that this is a \textit{blended} problem which combines DS and ML concepts. Specifically, the correlation challenge is DS, while the imbalanced datasets and choice of classifier challenges are ML-specific.

\paragraph{Solution Workflow Layer}

The case study implicitly defines a workflow through its use of Python scripts\footnote{\url{https://github.com/moosavianmz/DetectingDepression}}. However, Mousavian \textit{et al.} represent the workflow as explicit blocks in the article~\cite{Mousavian2021}. Therefore it is straightforward to classify each task in the workflow as DS or ML as done for the other case studies.

Table~\ref{tab:cs_nipype} shows the workflow as defined in the article of Mousavian \textit{et al.}. The steps present in the article clearly identify that the majority of the workflow is \textit{DS}. In particular, only the feature selection and actual classification steps are ML.

\paragraph{Implementation Layer}

As the workflow of this case study is mostly DS, it follows that the implementation is also very \textit{DS}. In particular, heavy use of DS libraries and tools are used such as \textit{dcm2niix} for format conversion, \textit{PyDeface} for removing facial structure from the images, and the fMRIPrep implementation itself\footnote{\url{https://github.com/nipreps/fmriprep}}.

%\paragraph{Remarks}

\subsubsection{CS\_Orange}
\label{sec:cs_orange}

The case study for the \textit{Orange} tool focuses on data mining analysis for Internet traffic from a smart school~\cite{Adekitan2019}.

% DSPS (maybe to MLPS) to MLWF/Blended to BI

\paragraph{Problem Layer}

The article from Adekitan \textit{et al.} is an exploratory analysis on using machine learning techniques for prediction of Internet traffic at an educational institution. The specific problem is to predict a classification for both download traffic and upload traffic among \textit{low}, \textit{slight}, \textit{moderate}, and \textit{heavy} data traffic. The input information is a numerical day, week, and month, along with the previous day's traffic and an average of the previous two days.

Multiple ML classifiers are used and compared in this analysis from both the Orange and KNIME tools mentioned in Section~\ref{sec:tools_graphical}. As the intention was to compare ML classifiers of two different workflow tools, we classify this problem as a \textit{blended} problem.

\paragraph{Solution Workflow Layer}

\begin{figure*}[tbh]
    \centering
    \includegraphics[width=0.8\textwidth]{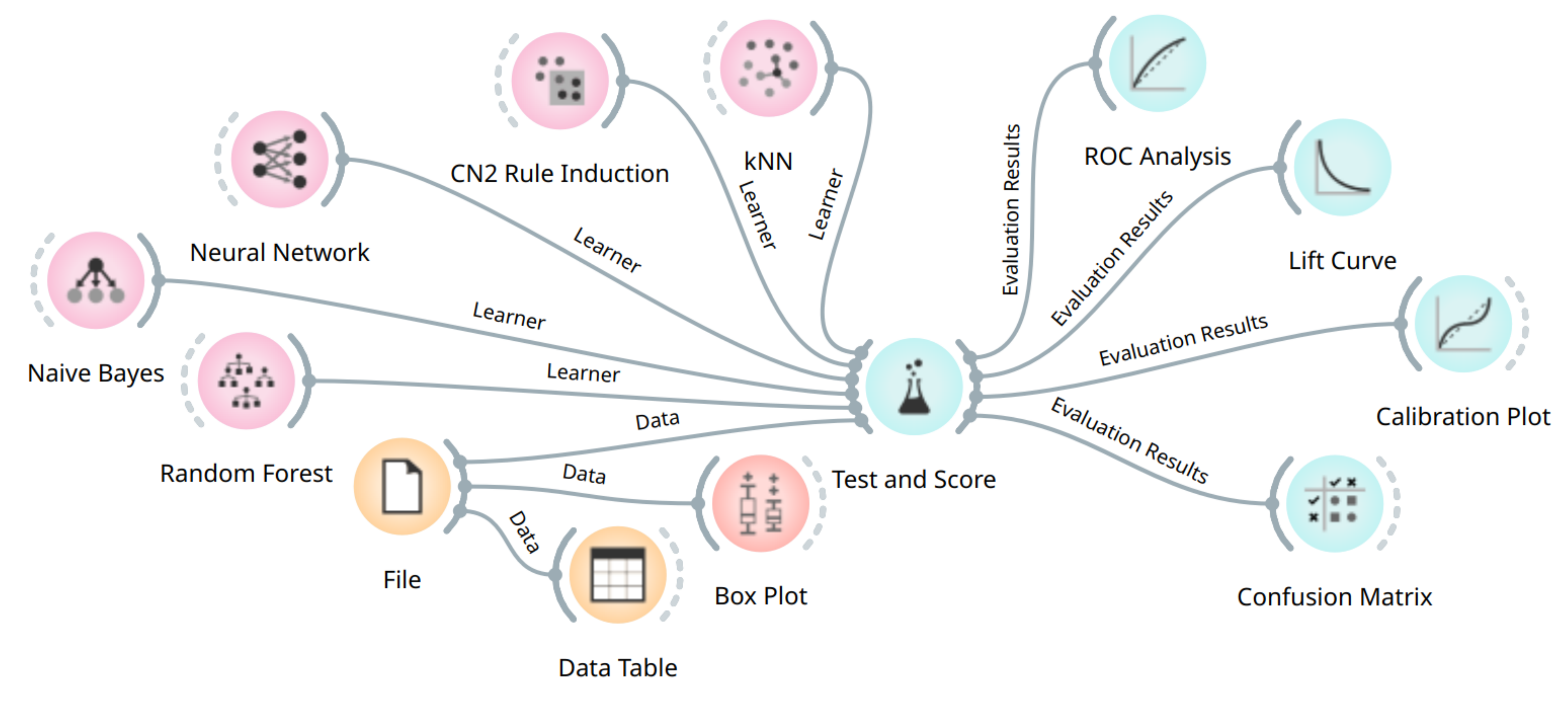}
    \caption{Solution workflow in the Orange tool. Adapted from~\cite{Adekitan2019}.}
    \label{fig:adekitan}
\end{figure*}

Figure~\ref{fig:adekitan} shows the workflow of Adekitan \textit{et al.} in the Orange tool, while the workflow for KNIME is found in their article~\cite{Adekitan2019}. The three \textit{general} components at the bottom (\textit{File}, \textit{Data Table}, and \textit{Box Plot}) are used to load the data and visualise it. The \textit{Test \& Score} component takes the five ML learners and the loaded data, and performs the ML classification task. The results are then passed to the four evaluation components on the right. We classify the \textit{Test \& Score} component, the learners, and the evaluation components as all ML components. There are no DS components in this workflow, however the feature engineering described in the article means that domain knowledge concerning the academic calendar of the institution has been encoded into the source data. Therefore, this workflow can be classified as mostly \textit{ML-based} with a DS feature engineering step.

\paragraph{Implementation Layer}

The execution for the Orange tool can be performed within the editor itself, or by calling the underlying Python code defined for each component\footnote{
Orange is currently unable to generate orchestration Python code from a workflow. See \url{https://github.com/biolab/orange3/issues/1341}}. In Orange, the code for the components used is built upon the scipy and scikit-learn modules. Therefore we classify this implementation as mostly \textit{ML-based}.

\paragraph{Remarks}
This case study represents an exploratory usage of ML techniques within a workflow. The authors performed some DS feature extraction on the data and then applied different classifiers to determine the performance. The classification performance found was quite low (55 to 63\% accuracy), indicating that further DS feature extraction may be required to improve classification performance.

% The authors did perform DS feature engineering to extract the days of the week and other time-based information, as this is relevant to the academic calendar of the institution. However, this DS processing step is applied outside of the workflow in both the Orange and KNIME tools, and is not submitting as accompanying materials in their article.

\subsubsection{CS\_Galaxy}
\label{sec:cs_galaxy}

% DSPS to DSWF/BWF to DSI/MLI/BI

The last case study for this report details a workflow for the Galaxy framework. The article of F{\"o}ll \textit{et al.} tackles supervised classification of urothelial (bladder) cancer using \textit{mass spectrometry imaging} (MSI)~\cite{Foell2021}.

\paragraph{Problem Layer}

The input data studied by F{\"o}ll \textit{et al.} is obtained using MSI. This imaging technique is performed on a slice of tissue. For each region in the sample, the instrument provides a spectrum of the masses of present biomolecules. That is, for each ``pixel'' of the sample image a one-dimensional spectrum is created where peaks correspond to a particular biocompound. This can be used to visualise and classify regions of the sample where a particular biomolecule is present.

In the problem of F{\"o}ll \textit{et al.}, this MSI data is manually labelled by an expert as containing either \textit{tumor} tissue or \textit{stroma} (connecting tissue). Tumor tissue is then further classified into \textit{invasive} or \textit{non-invasive}. Therefore the problem statement is to develop a classification workflow which can pre-process and classify this unique spectral data. This problem can be said to be \textit{DS} as it does not refer to the classification techniques used.

\paragraph{Solution Workflow Layer}

An extract from the Galaxy workflow of F{\"o}ll \textit{et al.} is seen in Figure~\ref{fig:galaxy_workflow} on page~\pageref{fig:galaxy_workflow}. In particular, these components are in the workflow which classifies tissue as a tumor or stroma. The \textit{MSI classification} component on the right-hand side of Figure~\ref{fig:galaxy_workflow} takes the MSI data and parameters and outputs a classification.

Similar to the other case studies, we present a broad analysis of the case study workflows\footnote{\url{https://github.com/foellmelanie/Bladder_MSI_Manuscript_Galaxy_links}} in Table~\ref{tab:cs_galaxy_workflow}. For each workflow created by the authors, we classify each component within as \textit{general}, \textit{domain-specific without ML concepts}, or \textit{domain-specific with ML concepts}\footnote{Note that file reading and writing were counted as general components, which somewhat inflates their number. As well, in WF3 and WF4 repeated components for splitting a dataset ten times were counted only once to avoid over-representation.}. A percentage of the components which are DS is reported in a column on the right-hand side of Table~\ref{tab:cs_galaxy_workflow} along with a classification of the workflow.

\begin{table*}[tbp]
    \centering
    \begin{tabular}{p{7cm}|c|c|c|c|c}
 & \textbf{Num. General} & \textbf{Num. DS} & \textbf{Num. DS} &  \textbf{\% DS} & \textbf{Classification}\\
 & \textbf{Comps.} & \textbf{Comps.} & \textbf{Comps.} &  &  \\
 \textbf{Workflow Name}&  & \textbf{(non-ML)} & \textbf{(ML)} &  &  \\\hline

WF1: Co-registration and ROI Extraction & 22 & 6 & 0 & 21 & Mod. DS\\
WF2: Data Handling and Preprocessing & 10 & 22 & 0 & 69 & DS\\
WF3: Classification tumor vs. stroma & 17 & 2 & 3 & 23 & Mod. DS\\
WF4: Classification Infiltrating vs. Non-infiltrating & 15 & 3 & 4 & 32 & Mod. DS \\
WF5: Visualization & 11 & 13 & 0 & 54 & DS\\
WF6: Annotating Potential Identities & 11 & 0 & 0 & 0 & General\\
        
    \end{tabular}
    \caption{Classification of Galaxy workflows for analysing urothelial cancer (F{\"o}ll \textit{et al.})~\cite{Foell2021}}
    \label{tab:cs_galaxy_workflow}
\end{table*}

%List of DS nodes:
% WF1:
% Teachmark regisration
% Coords of ROI
% MSI mz images
% Overlay
% Projective transformation
% Switch axis coordinates 

% WF2 (general nodes)
% 6 x Files
% 2 x Filter
% 2 x Build List

% WF3 
% 17 General (collapsing datasets)
% 5 DS

% WF4
% 15 General
% 7 DS

% WF5
% 11 General
% 13 DS

Table~\ref{tab:cs_galaxy_workflow} indicates that the workflows for this case study range from moderately DS to strongly DS. General components are used for dataset loading and manipulation while the DS components perform the non-trivial work. There are no non-domain-specific ML components in these workflows, and the component \textit{MSI Classification} performs the domain-specific classification computations. Thus it is clear that this is a \textit{DS} workflow.

\paragraph{Implementation Layer}

Workflows are run by Galaxy through the web-based tool on either a public or private server. The individual components are defined through XML wrappers which define how to run the underlying tool.

For these particular workflows, the majority of the DS components are specific to MSI as they have been created by the authors in a previous work~\cite{Foell2019}. These components are wrappers around the Cardinal tool, which is a R language module specifically for analysing mass spectrometry-based imaging~\cite{Bemis2015}. Thus the implementation of this workflow is mostly \textit{DS}.

% \paragraph{Remarks}

% Heavily geared towards genomics. Framework is very good though. To aid construction of DS workflow, offers many components such as downloading from DS databases, computational components. Implementation is performed through execution in the Galaxy client. Web-based and offers easy sharing and collaboration(?). Lots of available workflows.

% \cite{Schmidt2019}
% \url{https://github.com/HelmGroup/Galaxy_modification_calling}

\section{Discussion}
\label{sec:discussion}

This section provides discussion for the main research topic of this article: \textit{what are the ways in which domain experts can use workflow-based tools and techniques to to solve their domain-specific problems using machine learning}. For this discussion, we first present the benefits and drawbacks for structuring this research topic using our three-layer framework organised into two dimensions. Then we examine each of the challenge questions introduced in Section~\ref{sec:intro} and present what we see to be remaining research and integration challenges for the software engineering community.

\subsection{Benefits and Drawbacks of the Three-Layer Framework}

This section discusses some benefits and drawbacks of organising this research topic as a three-layer framework with inter- and intra-layer transformations.

\subsubsection{Benefits}

\paragraph{Separation of Concerns}

The main benefit of our framework is the separation of concerns into the three layers: \textit{problem}, \textit{workflow}, and \textit{implementation}. Similar to domain-specific languages, this ensures that the domain expert first addresses the \textit{problem space} which they are familiar with, rather than dealing with the accidental complexity of the workflow and implementation spaces. The framework defines transformations between these layers, offering the domain expert a structured way of progressing their solution.

%Around the stack of the framework, we have the rest of the application. We isolate the essential complexity of the problem, design, implementation. We're observing that the design is simple because it is a workflow, and this is reflected in the machine learning world with the pipeline. Don't want to do things at implementation level. Want to stay up at workflow level for essential complexity.

This separation of concerns is also present in the workflow literature. For example, Lamprecht \textit{et al.}~\cite{Lamprecht2021} define six stages of workflows over time: \textit{question or hypothesis}, \textit{conceptual workflow}, \textit{abstract workflow} (sequences of tools but not fully configured), \textit{concrete workflow} (ready to run), \textit{production workflow} (ready for reuse), and \textit{scientific results}. The first two stages of \textit{question/hypothesis} and \textit{conceptual workflow} thus map onto our notion of domain-specific problem.

\paragraph{Implicit versus Explicit Workflows}

Explicit workflows are both conceptually (and literally) at the centre of our framework. This is because we see numerous benefits with this formalism for domain experts to use in combination with ML.

First, it is obvious that there is compatibility between the use of scientific workflows and ML pipelines. They share the same underlying formalism due to the same concept of control and data flow, as well as concerns about modularity and reuse. A workflow-based approach also seems to be very amenable to visualisation and manipulation in graphical tools, allowing non-experts to quickly build a workflow for their problem.

Second, these standalone components are a useful abstraction over the technical details and complexity of ML approaches. The domain expert does not have to become familiar with the ML libraries or in some cases even a programming language to orchestrate the workflow. Again, this is in concordance with the principles of domain-specific engineering where the domain expert should focus on manipulating concepts within their domain.

Third, providing explicit components to the domain expert allows for enhanced traceability, reuse, scientific replication. As seen with the Galaxy tool, input and output history can be kept for every component in a workflow, along with explicit versioning and supporting information.

Fourth, a standardised workflow system can offer enhanced benefits for deployment on cloud or high-performance systems. For example, Lehmann \textit{et al.} discuss the scalability benefits gained when porting an implicit workflow orchestrated with the Bash shell language to the Nextflow workflow system~\cite{Lehmann2021}.

In some domains, the use of explicit workflows is a best practice. For example, Poldrack \textit{et al.} discuss the use of nipype for reproducible and scalable workflows in neuroscience~\cite{Poldrack2018}, while Reiter \textit{et al.} provide a detailed article of techniques for biology experts to get started with workflows\cite{Reiter2021}. However, other fields may not have such a strong culture of workflows and still recommend coding for problem solving. As an illustrative example, Wang \textit{et al.} recently suggest for material scientists to use PyTorch in Python for ML purposes~\cite{Wang2020}.

%\cite{Freeman2015} Argues against using MATLAB in neuroscience

\paragraph{Focus on Transformations}

Another benefit of our framework is the focus on transformations between regions of layers, as well as between the layers themselves. These transformations can be phrased as challenge questions which are relevant to both software engineers and the domain experts who must use these transformations. This provides a clear intent for each transformation and allows for analysis and identification of current approaches and new techniques. %This is further discussed in Section~\ref{sec:discuss_challenges} where current research challenges are linked to these challenge questions.

\subsubsection{Drawbacks}

There are a number of drawbacks to our organisation of the research problem onto this three-layer framework.

The first is that the two dimensions selected of \textit{domain-specific} and \textit{employing machine learning} are not orthogonal as mentioned in Section~\ref{sec:dimensions}. The divisions of these dimensions into regions is also crude as we are unable to provide specific metrics to divide problems, workflows, or implementations due to the qualitative nature of these dimensions. In particular, the classifications of the problem, workflows, and implementations for the case studies in Section~\ref{sec:case_studies} are very \textit{ad-hoc}.

Second, restricting these complex systems onto three layers is a gross simplification. In particular, we acknowledge that the implementation layer is most likely made up of numerous layers of domain-specific or general programming languages. We have classified implementation code in some case studies as \textit{domain-specific} when this is just the top layer of what may be \textit{machine learning} or \textit{general} code at the lowest layer.

Lastly, we also acknowledge the incompleteness of this article to cover the research topic. It is impossible to fairly cover all domains or to give an impression of how prevalent the usage of any tool or technique is within a domain. This lack of comprehensiveness may render our framework less applicable when applied to a particular domain.

\subsection{Challenges and Future Research Directions}
\label{sec:discuss_challenges}

In this section, we again present the challenge questions from the introduction in Section~\ref{sec:intro}. For each question we then present our thoughts on how this challenge question has been addressed by the tools and case studies seen in this article. We present the challenges and potential research directions for each question.

\subsubsection{Mapping a DS problem to a form suitable for ML}

The first challenge we have selected focuses on the \textit{problem layer}. That is, how to assist the domain expert to choose the machine learning techniques which may assist them.

Our analysis in Section~\ref{sec:tools_framework_relation} indicates that none of the workflow frameworks discussed in this article tackle this challenge. This may be expected as the challenge is at the problem level, not the workflow layer.

This challenge is however very relevant to a domain expert. For example, most of the case studies examined in Section~\ref{sec:case_studies} directly define a ML or blended problem. This could indicate that domain experts are having to gain sufficient machine learning knowledge to begin their study, instead of leaving the problem as a DS problem. In particular, the case study of Kaggle (Section~\ref{sec:cs_kaggle}) shows that domain experts will go to great lengths to obtain ML expertise on their problem.

Addressing this challenge involves bringing together semantic information from the DS and ML domains. In particular, ontological information could be used to match problems in a particular domain with a ML specification~\cite{Lamprecht2021}.

\subsubsection{Providing a solution workflow for a DS and/or ML problem}

The second challenge we observe is for the domain expert to efficiently discover and/or build a workflow which solves their problem.

Many frameworks use a repository approach to improve the discoverability of workflows. That is, they provide a website for domain experts to search for a workflow which suits their needs~\footnote{Examples include \url{https://workflowhub.eu/}, \url{httpsL//hub.knime.com}, and \url{https://nf-co.re}.}. We also point to the impressive Collective Knowledge framework~\cite{Fursin2021} for a repository focusing on AI, ML, and system research\footnote{\url{https://cknowledge.io}}.

However there still remains a challenge to connect the workflow solutions present in the repository with the problem faced by the domain expert~\cite{Garijo2017}. The literature discusses manual and automatic semantic extraction techniques which can assist domain experts in finding workflows~\cite{Diaz2017,Paul-Gilloteaux2021}. However, enabling this at-scale across multiple domains and tools will continue to be a challenge.

As an example of the rich information to extract from workflows, we point to the work of Lamprecht \textit{et al.} who discuss automatic discovery and the static analysis of workflows~\cite{Lamprecht2021}. This includes technical parameters (versioning, FAIRness metrics, usage, etc.), domain-specific considerations (relevance of components to a domain, similarity to existing workflows, type and format of results, etc.), and community influence (citations, comments, ratings, etc.).

Automated \textit{workflow composition} can also assist domain experts in building their workflows. A promising approach is to combine the techniques of AutoML~\cite{He2021} with domain knowledge about required domain-specific components~\cite{Kasalica2020,Gil2019}.

Whether a workflow is found or not, the domain expert is likely to want to add their own components. Thus another sub-challenge is to improve the suggestion possibilities for domain experts. Recommendations are found in the KNIME, Galaxy, and low-code tools~\cite{Almonte2020,Kumar2021}, but we see further potential in this research area. For example, suggesting larger pieces of workflows, improved semantic reasoning such that components relate directly to the domain~\cite{McIver2015}, and employing machine learning techniques themselves to suggest components~\cite{Nouri2021}.

\subsubsection{Allowing the domain expert to experiment with appropriate ML components in a DS workflow}

An interesting challenge is to encourage and assist a domain expert with experimenting with ML techniques within a workflow. For example, the Orange tool (Section~\ref{sec:cs_orange} makes it simple to add ML components to a workflow and visualise the results. This sort of visual experimentation fits perfectly with the component-based nature of workflows and ties in well with the challenge of improving the automatic recommendation systems. This experimentation step can also be partially automated by integrating AutoML techniques~\cite{He2021} or recommender systems~\cite{Kumar2021} into the workflow tool to dynamically react to workflow changes.

\subsubsection{Adding DS knowledge to improve ML performance}

Another research challenge is how to utilise the DS knowledge of a domain expert to directly improve the performance of ML techniques. This is seen in some case studies (such as in Section~\ref{sec:cs_kaggle} and Section~\ref{sec:cs_nipype}) where the features themselves were modified to take into domain knowledge. As with other challenges described here, one approach may be to combine domain knowledge represented in an ontology with suggestions for features to extract, such as provided by unsupervised feature extraction~\cite{Solorio-Fernandez2020} or ontology embeddings~\cite{Kulmanov2021}.

\subsubsection{Producing an implementation from a workflow which is well-suited for a domain expert}

Once a domain expert has created their workflow they must be able to run it in a scalable manner. This is addressed in multiple tools from Section~\ref{sec:tools}, but there will always be further challenges to ensure that a domain expert can deploy their solution on the correct infrastructure. For example, code could be generated or parameterised based on the domain-specific tasks or data operated on, such as image- or voxel-based datasets~\cite{Deelman2019}. The rise of containerisation also opens up many challenges to ensure that these containers are distributed for optimal scalability and security~\cite{Poldrack2018}.

\subsubsection{Extracting a workflow from an existing implementation (code or notebook)}

The final challenge we highlight in our article is to convert the existing implementations a domain expert may have into workflows. Amongst other benefits, this would improve the modularisation and dissemination of these solutions~\cite{Carvalho2017}.

For example, Jupyter notebooks~\footnote{\url{https://jupyter.org/}} are a well-known paradigm for storage, dissemination, and reproduction of experimental results~\cite{Oakes2019}. Each \textit{cell} in a notebook contains text or executable code, where the results of code are shown directly underneath. This format thus provides a narrative to provide context for the code, which is useful for disseminating results or tutorials on a topic.

Rule \textit{et al.} suggest that scientists spend time to make Jupyter notebooks themselves form part of a workflow~\cite{Rule2019}. An interesting line of research is therefore to develop tooling and techniques to automate this process, such that the legacy notebooks of domain experts or machine learning experts\footnote{See Quaranta \textit{et al.} for a dataset of Kaggle notebooks~\cite{Quaranta2021}.} can be automatically promoted to explicit workflows~\cite{Carvalho2017}.

\subsubsection{External Challenges}

%Address problem space, how for domain experts to solve the right problem?

%\cite{Burgueno2019} Modelling body of knowledge

Beyond the challenges related to the framework itself, we also identify two other challenges which are important to increase the impact of addressing the problems specified in this article. These challenges are: a) strengthening the workflow community as a whole, and b) proposing tools and techniques to guide a domain expert in solving their problems.

\paragraph{Strengthening the Workflow Community}

For domain experts to be able to effectively use workflows to solve their problems using ML, there must be a strong cross-domain workflow community. This community will then be able to pool knowledge and resources to best solve domain problems.

The excellent article of da Silva \textit{et al.} suggests current challenges and proposed activities in a workflow community context~\cite{Silva2021}. The challenges they see are: \textit{FAIR computational workflows}, \textit{AI workflows}, \textit{exascale challenges and beyond}, \textit{APIs, reuse, interoperability and standards}, \textit{training and education}, and \textit{building a workflows community}.

We also see other avenues to strengthen the workflow community. In particular, we note the recent research and commercial interest of \textit{low-code} platforms which are in some cases workflow management tools~\cite{Ihirwe2020,Bock2021}. It may be possible to leverage this interest into further developing workflow management systems by providing support for commercial domains. For example, the KNIME tool (Section~\ref{sec:tools_graphical} started development for solving pharmaceutical applications, but has now evolved to offer a commercial solution.

Crowdsourcing knowledge is also another possibility to build up the workflow community. For example, Paul-Gilloteaux \textit{et al.} suggest the organisation of regular ``taggathons'' to annotate tools, workflows, components, databases, and training materials with terms from an ontology~\cite{Paul-Gilloteaux2021}.

We also point towards Kaggle (Section~\ref{sec:cs_kaggle}) as an interesting community of domain and ML experts. Despite the issues with crowdsourcing~\cite{Tauchert2020,Bumann2021}, it may be possible to further utilise this pool of knowledge. In particular, we suggest that offering incentives for competition participants to include explicit workflows and reusable components in their solution may assist with reuse of their efforts.

Bringing in further expertise from software engineering sub-fields could also bring benefits to the workflow community. In particular, we draw from our own expertise in model-driven engineering to suggest that there are many research avenues to explore.

For instance, the multi- view/formalism/level of abstraction approach of multi-paradigm modelling may assist in reducing the cognitive complexity of domain experts~\cite{Giese2006}. A concrete example is providing \textit{views} on a workflow such that the domain expert can focus on different aspects of the workflow as needed.

Another research avenue would be integration of model management approaches such as modelling variability and uncertainty techniques into workflow management tools~\cite{Famelis2019}. The last research avenue we propose would be the integration of verification and validity techniques such as recording performance metrics~\cite{David2018}, enhancing type safety~\cite{Evans2022}, and checking for formal properties such as reachability~\cite{Fabra2018}.

\subsubsection{Guiding the Domain Expert}

%\cite{Singh2019} details some tool considerations in this dimension.

The last challenge we mention in this article is how to guide the domain expert in both finding the best practices and tools for their domain, as well as their path in the framework.

Recently there have been articles in multiple domains walking a domain expert through the best tools and techniques available to employ ML~\cite{Huang2021,Marx2020,Wang2020,Lafuente2021,Ruf2021}. For example, Nakhle and Harfouche provide four detailed Jupyter notebooks\footnote{Available here: \url{https://github.com/HarfoucheLab/Ready-Steady-Go-AI}} walking domain experts in phenomics (plant sciences) through four steps of a ML task~\cite{Nakhle2021}.

These four steps (\textit{[image] dataset selection}, \textit{data preprocessing}, \textit{data analysis}, and \textit{performance analysis and explanation}) are representative of most ML workflows. Therefore we suggest that similar dissemination efforts in different domains may assist domain experts. In particular, collaborative knowledge bases for a domain expert to navigate the tools and resources available in their domain may be useful.

%Don’t reinvent wheel. Take best of practice like Galaxy and disseminate to other fields. Answer the question: Does anything need to be added to this framework/tools to support other domains/machine learning?

Another tooling effort could be to dynamically assist the domain expert in producing template workflows based on their domain-specific problem~\cite{Lee2020,Nalchigar2020}. However, this raises the question of how to assist the domain expert through the regions of our framework in Section~\ref{sec:overview}.

For example, consider three approaches to take a \textit{DS} problem and arrive at a \textit{blended} workflow. The first approach is on the problem level, where the domain expert is provided with basic ML knowledge to assist them in refining the DS problem to include ML concepts. The second approach is to immediately build a workflow, and then use AutoML~\cite{He2021} techniques, assist the user in experimentation (as in the Orange tool), or use ontological recommendations~\cite{McIver2015} to complete the workflow. The last approach is to follow principles from \textit{human-guided machine learning} to iteratively build out the workflow~\cite{Santos2019,Gil2019,DOrazio2019}. In these three approaches, more or less automation may be appropriate depending on the task and user~\cite{Wang2021}.

A further consideration is whether to hide or expose the ML concepts and components based on the ML knowledge of the domain expert. This could allow a user to work with a mostly \textit{DS} workflow, and over time adjust the workflow towards a \textit{blended} workflow as they gain insight and familiarity with the ML components. 

\section{Conclusion}
\label{sec:conclusion}

This article presents a conceptual framework to structure the process and tools whereby domain experts can utilise machine learning to solve their problems. In particular, we focus on the computational workflow representation of solutions where executable components are connected by control and data flow edges. Examining the state-of-the-practice, we identify six key challenges that a domain expert may face in developing an executable workflow:
\begin{itemize}
    \item Map a DS problem to a form suitable for ML
    \item Obtain a solution workflow for a DS and/or ML problem
    \item Experiment with ML tools and techniques within a workflow
    \item Add DS knowledge to improve ML performance (e.g., feature engineering)
    \item Produce an implementation from a workflow which is well-suited for a domain expert (in terms of scalability, DS tooling, etc.)
    \item Extract a workflow from an existing implementation (code, Jupyter notebook)
\end{itemize}

These challenges are represented by transformation within regions of our conceptual framework. This framework has three layers, consisting of the \textit{problem layer}, \textit{workflow solution layer}, and \textit{implementation layer}. Each layer is further structured with two dimensions representing the \textit{domain specificity} and \textit{machine learning usage} of the artefacts on that layer.

This conceptual framework structures our investigation of the state-of-the-practice of how domain experts are employing machine learning. In particular, a selection of textual and graphical workflow tools are presented to illustrate tool support for the challenges we have identified. Case studies selected from recent works in various domains further explore how the problems, workflows, and implementations created by domain experts are heterogeneous in terms of the amount of domain specificity and machine learning usage. We also provide a short discussion on each challenge to indicate possible research directions.

This article thus forms a basis for further discussion and research into assisting domain experts with developing workflow solutions which employ machine learning. Integrating best practices from software engineering and across tools will reduce the friction for domain experts to utilise these powerful techniques and unlock new possibilities in their application to pressing scientific issues.

\begin{acknowledgements}
The authors would like to thank our colleagues Jessie Galasso-Carbonnel and Istv\'{a}n D\'{a}vid for their insightful discussions on this article.
\end{acknowledgements}

\section*{Conflict of interest}

The authors declare that they have no conflict of interest.

\bibliographystyle{spmpsci}      % mathematics and physical sciences
\bibliography{dsml-workflows}   % name your BibTeX data base

\end{document}